\documentclass[%
 aip,
 amsmath,amssymb,
 reprint,%
]{revtex4-1}

\usepackage[draft]{minted}
\usepackage{nameref}
\usepackage[dvips]{graphicx}
\usepackage[utf8]{inputenc}
\usepackage{amsmath,amssymb,latexsym,MnSymbol,bm}
\usepackage{enumitem}

\usepackage[table]{xcolor}
\usepackage{xcolor}
\usepackage[%
  colorlinks=true,
  urlcolor=blue,
  linkcolor=blue,
  citecolor=blue
]{hyperref}

\newcommand{\ordpy}{\texttt{ordpy}}
\newcommand{\numpy}{\texttt{numpy}}
\newcommand{\scipy}{\texttt{scipy}}
\newcommand{\networkx}{\texttt{networkx}}
\newcommand{\igraph}{\texttt{igraph}}
\newcommand{\graphtool}{\texttt{graph\_tool}}

\newcommand{\tisean}{\texttt{tisean}}
\newcommand{\powerlaw}{\texttt{powerlaw}}
\newcommand{\pyunicorn}{\texttt{pyunicorn}}

\newcommand{\symbolicsequence}{\texttt{ordinal\_sequence}}
\newcommand{\symbolicdistribution}{\texttt{ordinal\_distribution}}
\newcommand{\missingstates}{\texttt{missing\_patterns}}
\newcommand{\permutationentropy}{\texttt{permutation\_entropy}}
\newcommand{\complexityentropy}{\texttt{complexity\_entropy}}
\newcommand{\maxcurveentropy}{\texttt{maximum\_complexity\_entropy}}
\newcommand{\mincurveentropy}{\texttt{minimum\_complexity\_entropy}}
\newcommand{\ordinalnetwork}{\texttt{ordinal\_network}}
\newcommand{\globalnodeentropy}{\texttt{global\_node\_entropy}}
\newcommand{\randomordinalnetwork}{\texttt{random\_ordinal\_network}}
\newcommand{\missinglinks}{\texttt{missing\_links}}
\newcommand{\tsallisentropy}{\texttt{tsallis\_entropy}}
\newcommand{\renyientropy}{\texttt{renyi\_entropy}}
\newcommand{\tsallis}{\texttt{tsallis\_complexity\_entropy}}
\newcommand{\renyi}{\texttt{renyi\_complexity\_entropy}}

\newcommand{\tieprecision}{\texttt{tie\_precision}}

\begin{document}

\title{ordpy: A Python package for data analysis with permutation entropy and ordinal network methods}

\author{Arthur\ A.\ B.\ Pessa}
\email{arthur\_pessa@hotmail.com}
\affiliation{Departamento de F\'isica, Universidade Estadual de Maring\'a -- Maring\'a, PR 87020-900, Brazil}

\author{Haroldo\ V.\ Ribeiro}
\email{hvr@dfi.uem.br}
\affiliation{Departamento de F\'isica, Universidade Estadual de Maring\'a -- Maring\'a, PR 87020-900, Brazil}

\date{\today}

\begin{abstract}
Since Bandt and Pompe's seminal work, permutation entropy has been used in several applications and is now an essential tool for time series analysis. Beyond becoming a popular and successful technique, permutation entropy inspired a framework for mapping time series into symbolic sequences that triggered the development of many other tools, including an approach for creating networks from time series known as ordinal networks. Despite the increasing popularity, the computational development of these methods is fragmented, and there were still no efforts focusing on creating a unified software package. Here we present \href{http://github.com/arthurpessa/ordpy}{\ordpy{}}, a simple and open-source Python module that implements permutation entropy and several of the principal methods related to Bandt and Pompe's framework to analyze time series and two-dimensional data. In particular, \ordpy{} implements permutation entropy, Tsallis and R\'enyi permutation entropies, complexity-entropy plane, complexity-entropy curves, missing ordinal patterns, ordinal networks, and missing ordinal transitions for one-dimensional (time series) and two-dimensional (images) data as well as their multiscale generalizations. We review some theoretical aspects of these tools and illustrate the use of \ordpy{} by replicating several literature results.
\end{abstract}

\maketitle

\begin{quotation}
Permutation entropy is a complexity measure and data analysis tool stemming from nonlinear time series analysis and information theory. In the almost two decades since its conception, this method has gained prominence and become extensively studied and used by researchers from several fields. The concept of ordinal patterns introduced with permutation entropy has also inspired a whole ecosystem of related techniques, including an approach to map time series into networks known as ordinal networks. However, this ecosystem of tools still lacks a more comprehensive numerical implementation, limiting the further spreading of ordinal methods,  especially to fields with less tradition in developing computational tools. In this article, we present \texttt{ordpy} – an open-source Python package implementing several tools related to ordinal patterns for the analysis of time series and images. We present \texttt{ordpy}’s functionalities together with a review of all pertinent theoretical developments, replicate several literature results, and highlight possible developments of ordinal methods that can be explored with \texttt{ordpy}.
\end{quotation}

\section{Introduction}

Stemming from a combination of ideas from nonlinear time series analysis~\cite{kantz2004nonlinear, bradley2015nonlinear} and information theory~\cite{shannon1948mathematical}, permutation entropy was first introduced in 2002 by Bandt and Pompe~\cite{bandt2002permutation} as a simple, robust, and computationally efficient complexity measure for time series. This complexity measure is defined as the Shannon entropy of a probability distribution associated with ordinal patterns evaluated from partitions of a time series -- a procedure known as the Bandt-Pompe symbolization approach. Permutation entropy and its underlying symbolization approach have become increasingly popular among researchers working with time series analysis, leading to successful applications in fields as diverse as biomedical sciences~\cite{nicolaou2012detection}, econophysics~\cite{zunino2009forbidden}, physical sciences~\cite{garland2018anomaly}, and engineering~\cite{yan2012permutation}. The uses of permutation entropy also span a large variety of goals such as monitoring the dynamical regime of a system~\cite{yan2012permutation}, detecting anomalies in time series~\cite{garland2018anomaly}, characterizing time series data~\cite{nicolaou2012detection}, testing for serial independence~\cite{garcia2008nonparametric}, and are further documented in review articles by Zanin~\textit{et al.}~\cite{zanin2012permutation}, Riedl~\textit{et al.}~\cite{riedl2013practical}, Amig{\'o}~\textit{et al.}~\cite{amigo2015ordinal}, and Keller \textit{et al.}~\cite{keller2017permutation}.

Permutation entropy's success is not limited to its practical usage as this approach has inspired numerous time series analysis tools. Some of these related methods consider different quantifiers for the ordinal probability distribution~\cite{rosso2007distinguishing, zunino2008fractional, carpi2010missing, parlitz2012classifying, unakafov2014conditional, liang2015eeg, ruan2019ordinal, zunino2015permutation, bandt2017new, ribeiro2017characterizing, jauregui2018characterization}, generalize the Bandt-Pompe symbolization algorithm to evaluate ordinal structures on multiple temporal scales~\cite{aziz2005multiscale, zunino2010permutation, morabito2012multivariate, zunino2012distinguishing}, include signal amplitude information~\cite{fadlallah2013weighted, xia2016permutation, azami2016amplitude, chen2018weighted}, and account for equal values in time series~\cite{bian2012modified, cuesta2018patterns}. Other works have generalized permutation entropy and its ordinal approach to two-dimensional data such as images~\cite{ribeiro2012complexity, zunino2016discriminating}. The ordinal patterns underlying permutation entropy have also been used for mapping time series and images into networks known as ordinal networks~\cite{small2013complex, mccullough2015time, small2018ordinal, pessa2019characterizing, pessa2020mapping, borges2019learning, chagas2020characterization}.

The original version of permutation entropy and its various generalizations represent an essential and appealing framework for data analysis, especially when considering the increasing availability of large data sets~\cite{economist2010deluge} and the steady demand for reliable and computationally efficient methods for extracting meaningful information from these data sets~\cite{mattmann2013vision, blei2017science}. However, most methods emerging from Bandt and Pompe's seminal work lack freely available computational implementation, and the exceptions are limited to a single or very few approaches. Here we help to fill this gap by presenting \ordpy{} -- a simple and open-source Python module implementing several of the principal methods related to Bandt and Pompe's framework. This module has been designed to be easily set up and installed as its only dependency is \numpy{}~\cite{harris2020array}, a fundamental Python library implementing array objects and fast math functions that operate on these objects. Beyond our preferences, the Python programming language has been chosen for its widespread use in scientific computing~\cite{harris2020array} and extensive community support~\cite{perkel2015pickup}.

We present \ordpy{}'s functions and illustrate their usage along with a review of the pertinent theoretical developments of permutation entropy and its related ordinal methods. Our work alternates between the mathematical description of the different techniques and the presentation of functions and code snippets that implement these data analysis tools. We further use \ordpy{} to replicate several literature results. The source-code of \ordpy{} is freely available on its git repository (\url{github.com/arthurpessa/ordpy}) together with the documentation of all \ordpy{}'s functions (\url{arthurpessa.github.io/ordpy}). We can install \ordpy{} using the Python Package Index (PyPI) via:
\begin{minted}{shell}
 $ pip install ordpy
\end{minted}
We further provide the code and data for replicating all analyses presented in this article as a Jupyter notebook~\cite{shen2014interactive, kluyver2016jupyter} on \href{http://github.com/arthurpessa/ordpy}{\ordpy{}}'s website.

\section{An overview of ordinal distributions, permutation entropy, and complexity-entropy plane} \label{sec:permutation_methods}

We start by presenting a short review of Bandt and Pompe's seminal permutation entropy~\cite{bandt2002permutation}. As we have already mentioned, permutation entropy is the Shannon entropy of a probability distribution related to ordinal (or permutation) patterns evaluated using sliding partitions over a time series. This probability distribution is the so-called ordinal distribution or distribution of ordinal patterns, and the symbolization process used to estimate this distribution is the Bandt-Pompe approach. To describe this process, let us consider an arbitrary time series $\{x_t\}_{t=1,\dots,N_x}$. First, we divide this time series into $n_x = N_x - (d_x-1)\tau_x$ overlapping partitions comprised of $d_x>1$ observations separated by $\tau_x\geq1$ time units. For given values of $d_x$ and $\tau_x$, each data partition can be represented by 
\begin{equation}\label{eq:1dpartition}
    w_p =  (x_{p}, x_{p + \tau_x}, x_{p + 2\tau_x}, \dots, x_{p + (d_x - 2)\tau_x}, x_{p + (d_x - 1)\tau_x})\,,
\end{equation}
where $p = 1, \dots, n_x$ is the partition index. The parameters $d_x$ and $\tau_x$ are the only two parameters of the Bandt-Pompe method: $d_x$ is the embedding dimension~\cite{bandt2002permutation} and $\tau_x$ is the embedding delay~\cite{zunino2010permutation}. It is worth remarking that Bandt and Pompe's original proposal was restricted to $\tau_x=1$ (that is, data partitions comprised of consecutive time series elements), and the embedding delay was further introduced by Cao \textit{et al.}~\cite{cao2004detecting} and Zunino \textit{et al.}~\cite{zunino2010permutation}. As we shall see, the choices of $d_x$ and $\tau_x$ are important, and there is research exclusively devoted to determining optimal values for these parameters~\cite{riedl2013practical, cuestafrau2019embedded, myers2020automatic}.

Next, for each partition $w_p$, we evaluate the permutation $\pi_p = (r_0, r_1, \dots, r_{d_x-1})$ of the index numbers $(0, 1, \dots, d_x - 1)$ that sorts the elements of $w_p$ in ascending order, that is, the permutation of the index numbers defined by the inequality $x_{p + r_0} \leq x_{p + r_1} \leq \dots \leq x_{p + r_{d_x-1}}$. In case of equal values, we maintain the occurrence order of the partition elements, that is, if $x_{p+r_{k-1}} = x_{p+r_k}$ then $r_{k-1} < r_{k}$ for $k=1,\dots,d_x-1$~\cite{cao2004detecting}. As an illustration, suppose we have $x_t = (5,3,2,2,7,9)$ and set $d_x = 4$ and $\tau_x = 1$. The first partition is $w_1 = (5,3,2,2)$, and sorting its elements we find $2 \leq 2 < 3 < 5$ or $x_{1+2} \leq x_{1+3} < x_{1+1} <x_{1+0}$. Thus, the permutation symbol (ordinal pattern) associated with $w_1$ is $\pi_1 = (2,3,1,0)$. 
Another possibility for dealing with ties among partition elements consists in adding a small noise perturbation without modifying all other ordering relations. This latter scheme was initially proposed in Bandt and Pompe's seminal work, but it is much less used in the literature. In most cases, time series data have enough resolution for making these equalities negligible; however, this issue can become critical for low-resolution signals~\cite{zunino2017permutation}.

\begin{figure*}[!ht]
\centering
\includegraphics[width=0.75\linewidth]{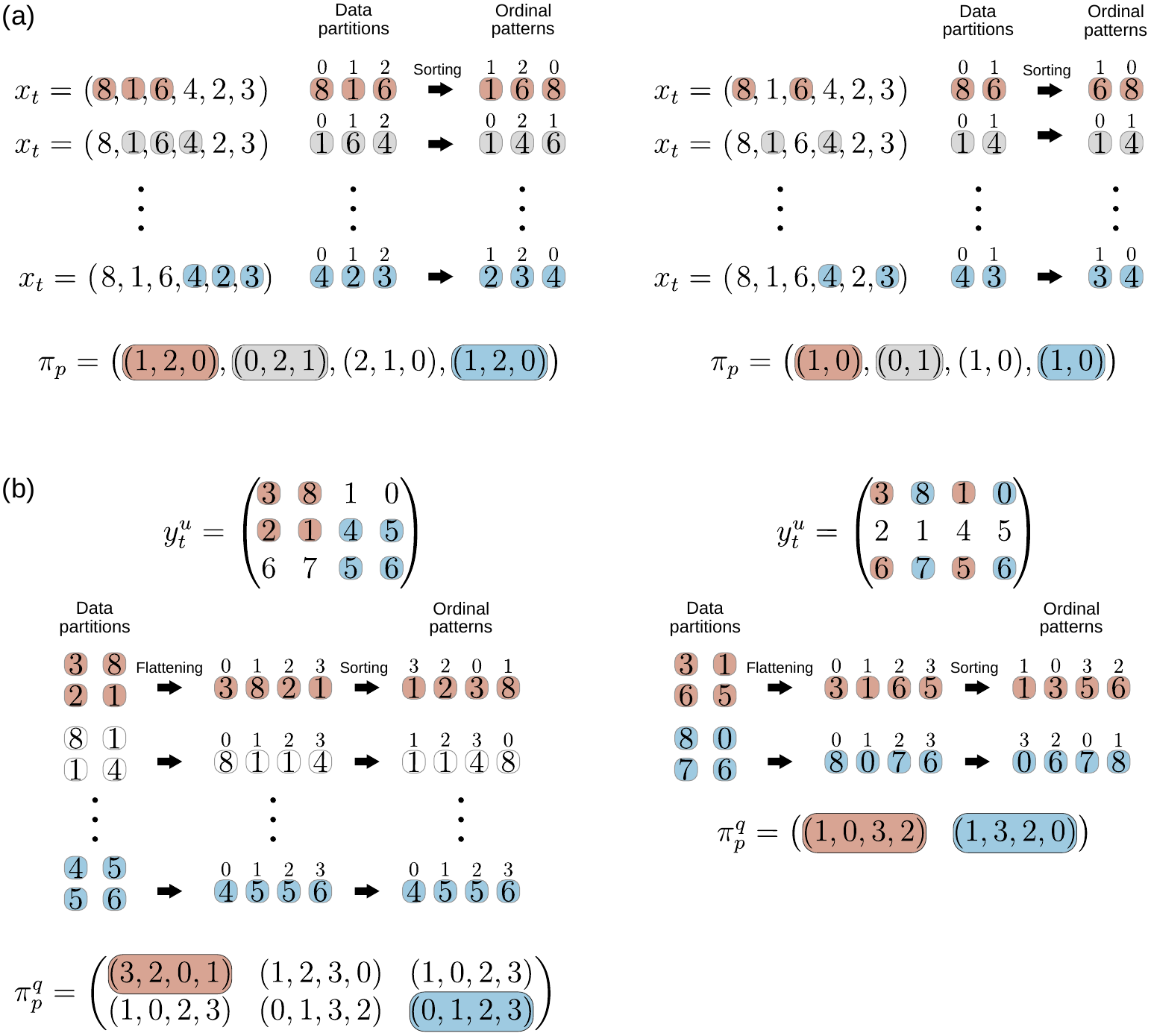}
\caption{The symbolization process of Bandt and Pompe. (a) Illustration of Bandt-Pompe method applied to a time series and the resulting ordinal sequences for embedding parameters $d_x=3$ and $\tau_x=1$ (left) and $d_x=2$ and $\tau_x=2$ (right). (b) Application of the two-dimensional version of the Bandt-Pompe method to a data array and the resulting ordinal sequences for embedding parameters $d_x=d_y=2$ and $\tau_x=\tau_1=1$ (left) and $d_x=d_y=2$ and $\tau_x=\tau_1=2$ (right). In both panels, the colored boxes indicate the data partitioning scheme for each set of embedding parameters, and the numbers above them represent the index numbers used for determining the ordinal patterns or permutation symbols.}
\label{fig:1}
\end{figure*}

After evaluating the permutation symbols associated with all data partitions, we obtain a symbolic sequence $\{\pi_p\}_{p = 1, \dots, n_x}$. The \ordpy's function \symbolicsequence{} returns this sequence as illustrated in the following code:
\begin{minted}{python}
>>> from ordpy import ordinal_sequence
>>> x = [5, 3, 2, 2, 7, 9]
>>> ordinal_sequence(x, dx=4, taux=1)
array([[2, 3, 1, 0],
       [1, 2, 0, 3],
       [0, 1, 2, 3]])
>>> ordinal_sequence([1.55, 1.54, 1.53], dx=2)
array([[1, 0], [1, 0]])
>>> ordinal_sequence([1.55, 1.54, 1.53], dx=2, 
... tie_precision=1)
array([[1, 0], [0, 1]])
\end{minted}
The last two examples illustrate the use of parameter \tieprecision{} that defines the number of decimals considered for establishing the ordinal relations. This parameter is available in most \ordpy{}'s functions and is particularly relevant when working with time series presenting equal values that could be mistaken by floating-point number representation. Figure~\ref{fig:1}(a) illustrates the application of the Bandt-Pompe approach for a simple time series and different values of $d_x$ and $\tau_x$.

The ordinal probability distribution $P = \{\rho_i(\Pi_i)\}_{i = 1, \dots, n_\pi}$ is simply the relative frequency of all possible permutations within the symbolic sequence, that is, 
\begin{equation}\label{eq:permutation_probability}
    \rho_i(\Pi_i) = \frac{\text{number of partitions of type} \ \Pi_i \ \text{in} \ 
    \{\pi_p\}}{n_x}\,,
\end{equation}
where $\Pi_i$ represents each one of the $n_\pi=d_x!$ different ordinal patterns. The following code shows how to obtain the ordinal distribution with \ordpy{}'s function \symbolicdistribution{}:
\begin{minted}{python}
>>> from ordpy import ordinal_distribution
>>> x = [5, 3, 2, 2, 7, 9]
>>> pis, rho = ordinal_distribution(x, dx=3)
>>> pis
array([[0, 1, 2],
       [1, 2, 0],
       [2, 1, 0]])
>>> rho
array([0.5 , 0.25, 0.25])
\end{minted}
The two arrays returned by \symbolicdistribution{} are the ordinal patterns and their corresponding relative frequencies, respectively. By default, \symbolicdistribution{} does not return non-occurring permutations (that is, those with $\rho_i(\Pi_i)=0$); however, the parameter \texttt{return\_missing} modifies this behavior as in:
\begin{minted}{python}
>>> from ordpy import ordinal_distribution
>>> x = [5, 3, 2, 2, 7, 9]
>>> pis, rho = ordinal_distribution(x, dx=3, 
... return_missing=True)
>>> pis
array([[0, 1, 2],
       [1, 2, 0],
       [2, 1, 0],
       [0, 2, 1],
       [1, 0, 2],
       [2, 0, 1]])
>>> rho
array([0.5 , 0.25, 0.25, 0.  , 0.  , 0.  ])
\end{minted}
Missing permutation symbols are always the latest elements of the returned array.

Having the ordinal probability distribution $P$, we can calculate its Shannon entropy~\cite{shannon1948mathematical} and define the permutation entropy as
\begin{equation}\label{eq:permutation_entropy}
    S(P) = -\sum_{i = 1}^{n_\pi} \rho_i(\Pi_{i})\log \rho_i(\Pi_{i})\,,
\end{equation}
where $\log(\dots)$ stands for the base-$2$ logarithm. Permutation entropy quantifies the randomness in the ordering dynamics of a time series such that $S\approx\log n_\pi$ indicates a random behavior, while $S\approx0$ implies a more regular dynamics. Because the maximum value of $S$ is $S_{\rm max} = \log{n_\pi}$, we can further define the normalized permutation entropy as
\begin{equation}\label{eq:normalized_pe}
    H(P) = \frac{S(P)}{\log{n_\pi}}\,,
\end{equation}
where the values of $H$ are restricted to the interval $[0,1]$. The \ordpy's function \permutationentropy{} calculates the values of $S$ and $H$ directly from a time series as illustrated in:
\begin{minted}{python}
>>> from ordpy import permutation_entropy
>>> x = [5, 3, 2, 2, 7, 9]
>>> permutation_entropy(x)
0.5802792108518123
>>> permutation_entropy(x, normalized=False, 
... base='e')
1.0397207708399179
\end{minted}
The \permutationentropy{} function uses the base-2 logarithm function by default; however, the parameter \texttt{base} can modify this behavior.

The embedding dimension $d_x$ defines the number of possible permutations $(n_\pi = d_x!)$, and following Bandt and Pompe's recommendation~\cite{bandt2002permutation}, it is common to choose the values of $d_x \in \{3, 4, 5, 6, 7\}$ to satisfy the condition $d_x! \ll N_x$ to obtain a reliable estimate of the ordinal probability distribution. Another less common choice is to use a value of $d_x$ such that $5 d_x! \leq N_x$~\cite{amigo2008combinatorial}. More recently, however, Cuesta-Frau \textit{et al.}~\cite{cuestafrau2019embedded} have shown that these requirements on $d_x$ can be considerably loosened in several situations related to classification tasks. The embedding delay $\tau_x$ defines a time scale for the system under analysis and is often set as $1$; however, different values of $\tau_x$ may inform about delayed feedback mechanisms and time-correlation structures. We present a more detailed discussion about the choices of $d_x$ and $\tau_x$ in Appendix~\ref{appendix:parameter}.

The permutation entropy framework was extended to two-dimensional data by Ribeiro \textit{et al.}~\cite{ribeiro2012complexity} and Zunino and Ribeiro~\cite{zunino2016discriminating}. To present this generalization, let us consider an arbitrary two-dimensional data array $\{y_t^u\}_{t = 1,\dots,N_x}^{u = 1,\dots,N_y}$ whose elements may represent pixels of an image. We further define the embedding dimensions $d_x$ and $d_y$ along the horizontal and vertical directions (respectively), and the corresponding embedding delays $\tau_x$ and $\tau_y$. Similarly to the one-dimensional case, we slice the data array in partitions of size $d_x \times d_y$ defined by 
\begin{equation}\label{eq:matrix_partition} 
    w_{p}^{q} = 
\begin{pmatrix}
    y_{p}^{q} & y_{p+\tau_x}^{q} & \dots & y_{p+(d_x-1)\tau_x}^{q} \\[.5em]
    y_{p}^{q+\tau_y} & y_{p+\tau_x}^{q+\tau_y}   & \dots & y_{p+(d_x-1)\tau_x}^{q+\tau_y} \\
    \vdots & \vdots & \ddots & \vdots\\
    y_{p}^{q+(d_x-1)\tau_x} & y_{p+\tau_x}^{q+(d_x-1)\tau_x}  & \dots & y_{p+(d_x-1)\tau_x}^{q+(d_y-1)\tau_y} \\
\end{pmatrix},
\end{equation}
where the indexes $p = 1,\dots, n_x$ and $q = 1,\dots, n_y$, with $n_x = N_x-(d_x-1)\tau_x$ and $n_y = N_y-(d_y-1)\tau_y$, cover all $n_x n_y$ data partitions. To associate a permutation symbol with each two-dimensional partition, we flatten the partitions $w_p^q$ line by line, that is,
\begin{equation}\label{eq:partition_index}
\begin{split}
    w_{p}^{q} =  & \left( y_{p}^{q}, y_{p+\tau_x}^{q}, \dots,
                    y_{p+(d_x-1)\tau_x}^{q}, \right.\\
                 & ~~y_{p}^{q+\tau_y}, y_{p+\tau_x}^{q+\tau_y}, \dots,         
                     y_{p+(d_x-1)\tau_x}^{q+\tau_y},\dots,\\
                 & \left. ~y_{p}^{q+(d_x-1)\tau_x}, y_{p+\tau_x}^{q+(d_x-1)\tau_x}, \dots, y_{p+(d_x-1)\tau_x}^{q+(d_y-1)\tau_y}\right)\,.
\end{split}
\end{equation}
As this procedure does not depend on a particular partition, we can simplify the notation by representing $w_{p}^{q}$ as
\begin{equation}\label{eq:partition}
    w_p^q = \left(\tilde{y}_0, \tilde{y}_1, \dots, \tilde{y}_{d_x d_y-2}, \tilde{y}_{d_x d_y-1}\right)\,,
\end{equation}
where $\tilde{y}_0 = y_{p}^{q},~\tilde{y}_1 = y_{p+\tau_x}^{q}$, and so on. Then, we evaluate the permutation symbol associated with each data partition as in the one-dimensional case to define the symbolic array $\{\pi_p^q\}_{p=1,\dots,n_x}^{q=1,\dots,n_y}$ related to the data set (Fig.~\ref{fig:1}{b} illustrates the Bandt-Pompe approach for two-dimensional data). From this array, we calculate the relative frequency for all $n_\pi = (d_x d_y)!$ possible permutations $\Pi_i$ via 
\begin{equation}\label{eq:permutation_probability_2d}
    \rho_i(\Pi_i) = \frac{\text{number of partitions of type $\Pi_i$ in } \{\pi_p^q\}}{n_x n_y}\,,
\end{equation}
where $i=1,\dots,n_\pi$, and so the ordinal probability distribution is $P = \{\rho_i(\Pi_i)\}_{i = 1, \dots, n_\pi}$. It is worth noticing that the ordering procedure defining the permutation symbols is no longer unique as in the one-dimensional case. For instance, we would find a different symbolic array by flattening the partitions $w_p^q$ column by column. However, different ordering procedures do not modify the set of elements comprising the ordinal probability distribution (only their order is changed)~\cite{ribeiro2012complexity}.

As in the one-dimensional case, the two-dimensional permutation entropy is simply the Shannon entropy of the ordinal distribution $P = \{\rho_i(\Pi_i)\}_{i = 1, \dots, n_\pi}$, so we can calculate the two-dimensional permutation entropy and its normalized version using Eqs.~\ref{eq:permutation_entropy} and \ref{eq:normalized_pe}, respectively. Only the total number of possible ordinal patterns ($n_\pi = (d_x d_y)!$ in the two-dimensional case) is modified.

Similarly to the one-dimensional case, the values of $d_x$ and $d_y$ are usually constrained by the condition $(d_x d_y)! \ll N_x N_y$ in order to obtain a reliable estimate of the ordinal distribution $P$~\cite{ribeiro2012complexity, zunino2016discriminating}. Naturally, this two-dimensional formulation recovers the one-dimensional case ($N_y=1$ for time series data) by setting $d_y = \tau_y = 1$. In \ordpy{}, the functions \symbolicsequence{}, \symbolicdistribution{} and \permutationentropy{} automatically implement this two-dimensional generalization when the input data is a two-dimensional array as in:
\begin{minted}{python}
>>> from ordpy import ordinal_sequence, 
... ordinal_distribution, permutation_entropy
>>> y = [[5, 3, 2], [2, 7, 9]]
>>> ordinal_sequence(y, dx=2, dy=2)
array([[[2, 1, 0, 3],
        [1, 0, 2, 3]]])
>>> ordinal_distribution(y, dx=2, dy=2)
(array([[1, 0, 2, 3],
        [2, 1, 0, 3]]), array([0.5, 0.5]))
>>> permutation_entropy(y, dx=2, dy=2)
0.21810429198553155
\end{minted}

In addition to permutation entropy, the complexity-entropy plane proposed by Rosso \textit{et al.}~\cite{rosso2007distinguishing} is another popular time series analysis tool directly related to Bandt and Pompe's symbolization approach. This method was initially introduced for distinguishing between chaotic and stochastic time series but has been successfully used as an effective discriminating tool in several other contexts~\cite{rosso2009detecting, zunino2010complexity, zunino2012efficiency, ribeiro2012songs, sigaki2019estimating}. The complexity-entropy plane combines the normalized permutation entropy $H$ (Eq.~\ref{eq:normalized_pe}) with an intensive statistical complexity measure $C$ (also calculated using the ordinal distribution) to build a two-dimensional representation space with the values of $C$ versus $H$. The statistical complexity $C$ used by Rosso \textit{et al.} is inspired by the work of Lopez-Ruiz \textit{et al.}~\cite{lopezruiz1995statistical} and is defined by the product of the normalized permutation and a normalized version of the Jensen-Shannon divergence~\cite{lin1991divergence} between the ordinal distribution $P = \{\rho_i(\Pi_i)\}_{i = 1, \dots, n_\pi}$ and the uniform distribution $U = \{ 1/n_\pi \}_{i = 1,\dots,n_\pi}$ (it is worth remembering that $n_\pi$ is the number of possible ordinal patterns). Mathematically, we can write this measure as
\begin{equation}~\label{eq:statistical_complexity}
    C(P) = \frac{D(P,U)H(P)}{D^{\rm max}}\,,
\end{equation}
where 
\begin{equation}
    D(P,U) = S[(P + U)/2] - \dfrac{1}{2}S(P) - \dfrac{1}{2}S(U) 
\end{equation}
is the Jensen-Shannon divergence and 
\begin{equation*} 
    D^{\rm max} = -\dfrac{1}{2}\left(\frac{n_\pi!+1}{n_\pi!}\log(n_\pi!+1)-2\log(2n_\pi!)+\log{n_\pi!}\right)
\end{equation*}
is a normalization constant. This latter constant expresses the maximum possible value of $D(P,U)$ occurring for $P = \{ \delta_{1,i} \}_{i = 1,\dots,n_\pi}$~\cite{lamberti2004intensive, martin2006generalized}, where
$\delta_{ij} =
\begin{cases}
1 & \text{if } i = j\\
0 & \text{if } i \neq j
\end{cases}$ 
is the Kronecker delta function.

Differently from permutation entropy, the statistical complexity $C$ is zero in both extremes of order (when only one permutation symbol occurs) and disorder (when all permutations are equally likely to happen). The value of  $C$ quantifies structural complexity and provides additional information that is not carried by the
value of $H$. Furthermore, $C$ is a nontrivial function of $H$ in the sense that for a given value of $H$, there exists a range of possible values for $C$~\cite{lamberti2004intensive, martin2006generalized, rosso2007distinguishing}. This happens because $H$ and $D$ are expressed by different sums of $\rho_i(\Pi_i)$ and there is thus no reason for assuming a univocal relationship between $H$ and $C$. 

To better illustrate this feature, let us assume (for simplicity) we replace the Jensen-Shannon divergence by the Euclidean distance between $P$ and $U$ (as in the seminal work of Lopez-Ruiz \textit{et al.}~\cite{lopezruiz1995statistical}), that is, $D(P,U) = \sum_{i=1}^{n_\pi} (\rho_i(\Pi_i)-1/n_\pi)^2$. In this case, the statistical complexity is
\begin{equation*}
    C(P) \propto -\left(\sum_{i=1}^{n_\pi} \rho_i(\Pi_i) \log \rho_i(\Pi_i)\right)\left(\sum_{i=1}^{n_\pi} (\rho_i(\Pi_i)-1/n_\pi)^2\right)\,,
\end{equation*}
and we can readily observe that different ordinal distributions $P=\{\rho_i(\Pi_i)\}_{i=1,\ldots,n_\pi}$ may lead to the same value of $H$ but different values of $C$ (or vice-versa). Let us further consider a particular ordinal distribution with three possible permutation symbols (this would be equivalent to having $d_x!=3$ or $(d_x d_y)!=3$, if possible), that is, $P=\{a,b,1-(a+b)\}$, where $a>0$ and $b>0$ are real numbers such that $(a+b)\leq1$ (to ensure the normalization of $P$). For this case, we have $S = - a \log a - b \log b - [1-(a+b)]\log[1-(a+b)]$ and $D = (a-1/3)^2 + (b-1/3)^2 + ([1-(a+b)]-1/3)^2$. Thus, for instance, if $a=0.79$ and $b=0.18$ or $a=0.80$ and $b=0.16$ we find the same value of $H = S/\log{3}\approx0.55$, but different values for $D$ ($0.32$ in the first case and $0.33$ in the second) and, consequently, for $C$. 

\begin{figure*}[!ht]
\centering
\includegraphics[width=0.8\linewidth]{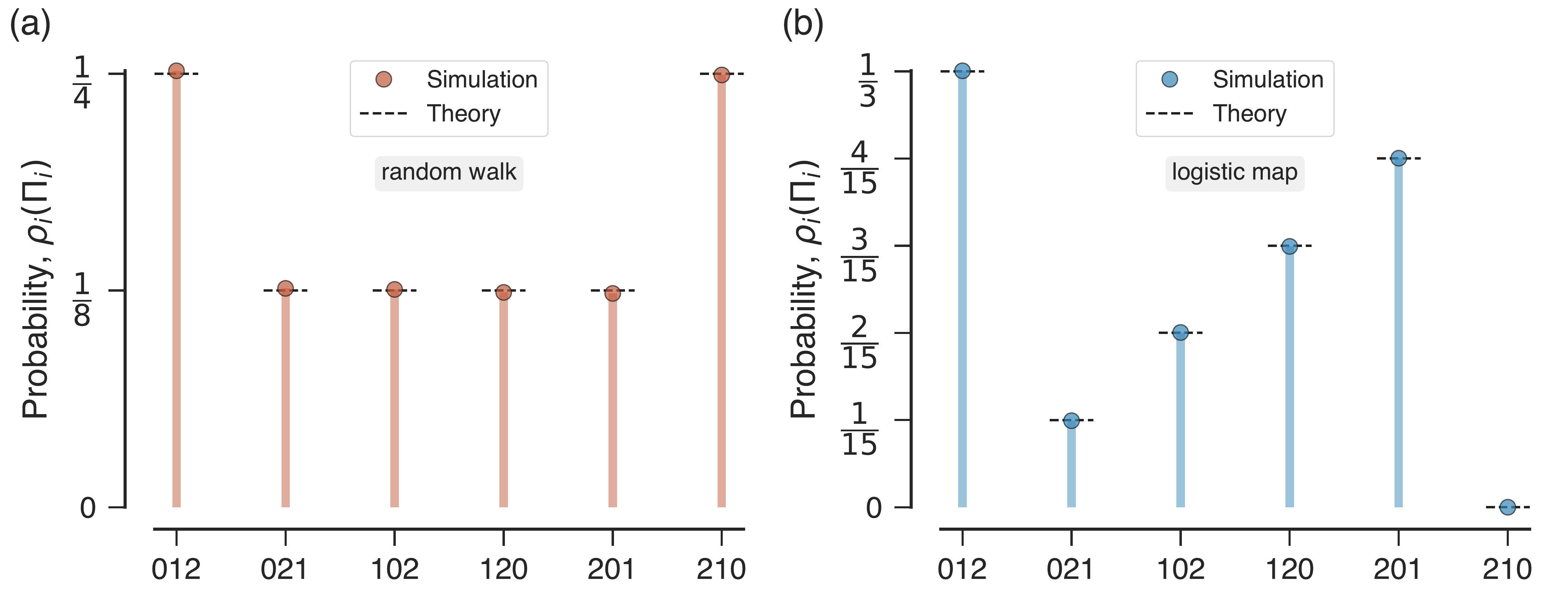}
\caption{Probability distributions of ordinal patterns for stochastic and deterministic series. (a) Comparison between the empirical probability distribution of ordinal patterns obtained from a simulated Gaussian random walk with $10^6$ steps and the exact distribution $P_{\rm walk}$ (dashed horizontal lines) for $d_x = 3$ and $\tau_x = 1$. (b) Comparison between the empirical probability distribution of ordinal patterns obtained from $10^6$ iterations of the logistic map at fully developed chaos and the exact distribution $P_{\rm logistic}$ (dashed horizontal lines) for $d_x = 3$ and $\tau_x = 1$. All results in this figure can be replicated by running a Jupyter notebook available at \href{http://github.com/arthurpessa/ordpy}{\ordpy{}}'s webpage.}
\label{fig:2}
\end{figure*}

In \ordpy{}, the \complexityentropy{} function simultaneously returns the values of $H$ and $C$ from time series as illustrated in:
\begin{minted}{python}
>>> from ordpy import complexity_entropy 
>>> complexity_entropy([4,7,9,10,6,11,3],
... dx=2)
(0.9182958340544894, 0.06112816548804511)
\end{minted}
Furthermore, the complexity-entropy plane was generalized for two-dimensional data~\cite{ribeiro2012complexity, zunino2016discriminating} (notice that the only changes are related to the process of estimating the ordinal distribution) and the \complexityentropy{} function also accepts two-dimensional arrays as input as shown in:
\begin{minted}{python}
>>> from ordpy import complexity_entropy
>>> complexity_entropy([[1,2,1],[8,3,4],
... [6,7,5]], dx=2, dy=2)
(0.3271564379782973, 0.2701200547320647)
\end{minted}

\section{Applications of Bandt and Pompe's framework with \ordpy{}}

This section presents more engaging applications of \ordpy{}'s functions by replicating literature results. We start by determining the ordinal probability distributions of two different time series of stochastic and chaotic nature, namely, a random walk with Gaussian steps and the logistic map at fully developed chaos (see Appendix~\ref{appendix:chaos} for definitions). We choose these two examples because their ordinal distributions are exactly known for some combinations of the embedding parameters~\cite{amigo2006order, bandt2007order}. More specifically, for $d_x = 3$ and $\tau_x = 1$, the probability distributions associated with the permutation symbols $\{(0,1,2), (0,2,1), (1,0,2), (1,2,0), (2,0,1), (2,1,0)\}$ are $P_{\rm walk} = \{{1}/{4}, {1}/{8}, {1}/{8}, {1}/{8}, {1}/{8}, {1}/{4}\}$ and $P_{\rm logistic} = \{{1}/{3}, {1}/{15}, {2}/{15}, {3}/{15}, {4}/{15}, 0\}$ for the random walk~\cite{bandt2007order} and the logistic map~\cite{amigo2006order}, respectively.

To numerically estimate these two ordinal distributions, we generate a time series from a Gaussian random walk process and another time series from iterations of the fully chaotic logistic map. In both cases, we have simulated one realization of each process with $10^6$ observations and used the \symbolicdistribution{} function. Figure~\ref{fig:2} shows that the exact ordinal distributions are in excellent agreement with simulated results obtained with \ordpy{}. It is intriguing to observe that the ordinal pattern $(2,1,0)$ (``descending permutation'') does not occur in the logistic series (it has probability zero). This fact is best understood as a feature directly associated with the intrinsic determinism of the logistic map dynamics~\cite{amigo2006order, amigo2007true}. As we shall discuss in the next section, investigations about such ``missing ordinal patterns'' are also useful for characterizing time series dynamics.

\begin{figure*}[!ht]
\centering
\includegraphics[width=1\linewidth]{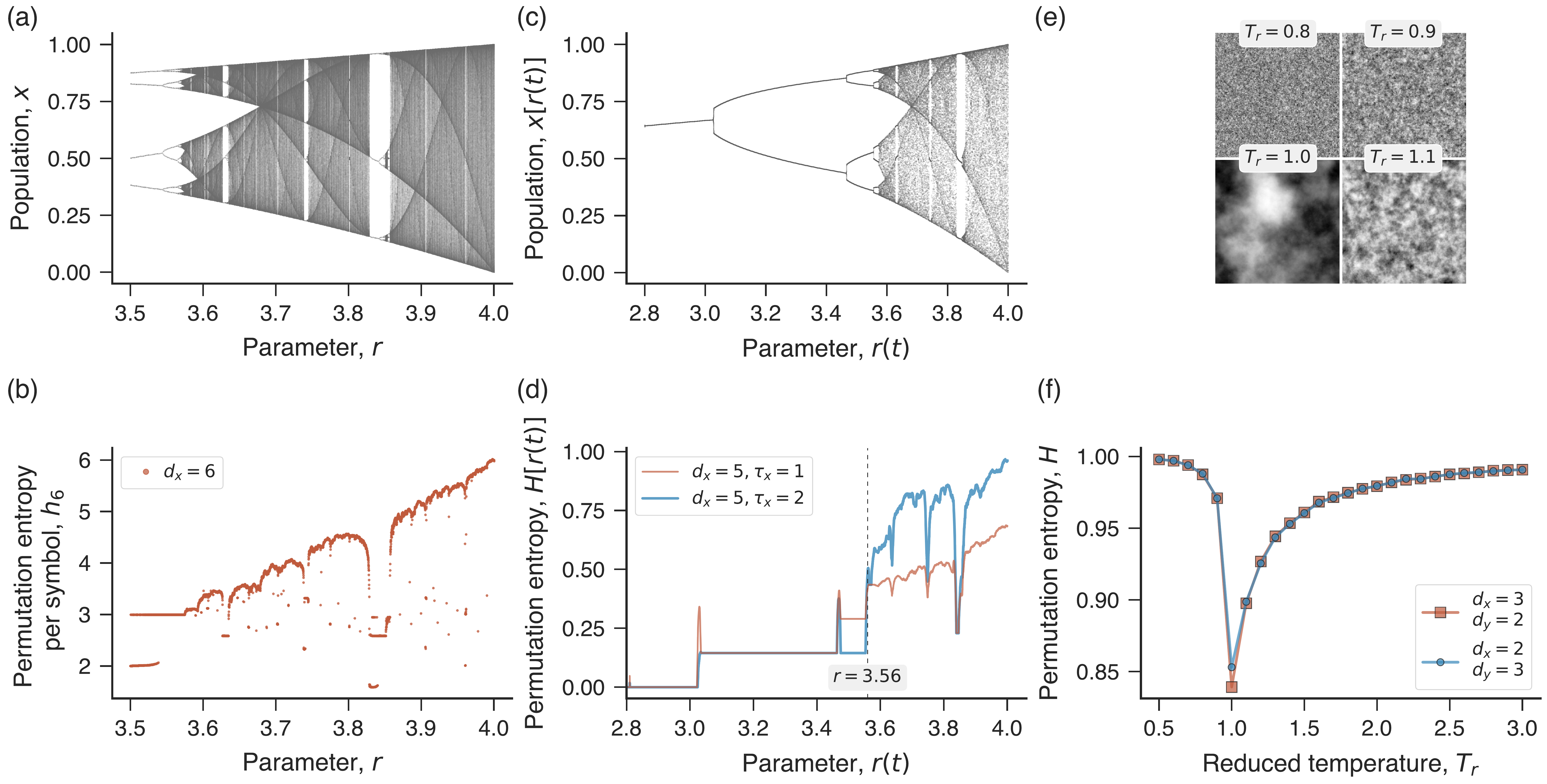}
\caption{Permutation entropy of one- and two-dimensional data. (a) Bifurcation diagram of the logistic map for $r$ between $3.5$ and $4$ in steps of size $10^{-4}$. (b) Permutation entropy per symbol of order six $(h_6)$ calculated from logistic time series with $10^6$ observations (random initial conditions) and $r \in \{3.5,3.5001,3.5002,\dots,4.0\}$. The embedding parameters are $d_x = 6$ and $\tau_x = 1$. (c) Time series of the transient logistic map obtained from the initial condition $x_0 = 0.65$, and by incrementing the logistic parameter $r$ at each iteration from $2.8$ to $4$ in steps of size $10^{-5}$. Despite appearing very similar to a bifurcation diagram, this result refers to a time series where each observation $x[r(t)]$ corresponds to a value $r(t)$. (d) Dependence of the normalized permutation entropy evaluated within a sliding window with 1024 observations of the original time series. Here $r(t)$ represents the logistic parameter at the end of each sliding window. The different curves show the results for $d_x = 5$ and $\tau_x = 1$ (red), and $d_x = 5$ and $\tau_x = 2$ (blue). The vertical line at $r = 3.56$ indicates the period-8 to period-16 bifurcation. (e) Ising surfaces obtained after $10^6$ Monte Carlo steps with reduced temperatures $T_r \in \{0.8, 0.9, 1.0, 1.1\}$. In these surfaces, dark gray shades indicate high lattice sites while light gray regions indicate the opposite. (f) Normalized permutation entropy as a function of the reduced temperature $T_r \in \{0.5,0.6,\dots,3.0\}$ for Ising surfaces of size $250 \times 250$ obtained after $10^6$ Monte Carlo steps. The different curves show the results for embedding parameters $d_x = 3$ and $d_y =2$ (red) and $d_x = 2$ and $d_y =3$ (blue), both with $\tau_x = \tau_y = 1$. All results in this figure can be reproduced by running a Jupyter notebook available at \href{http://github.com/arthurpessa/ordpy}{\ordpy{}}'s webpage.}
\label{fig:3}
\end{figure*}

To better illustrate the use of the \permutationentropy{} function, we partially reproduce Bandt and Pompe's analysis of the logistic map (Fig.~2 of Ref.~\onlinecite{bandt2002permutation}). We generate time series consisting of $10^6$ iterations of the logistic map for each value of parameter $r \in \{3.5, 3.5001, 3.5002, \dots, 4.0\}$ (see Appendix~\ref{appendix:chaos} for definitions). Next, we calculate the permutation entropy $S$ for each of these 5001 time series using \permutationentropy{} with embedding parameters $d_x = 6$ and $\tau_x = 1$. We further divide the permutation entropy by $5$ to obtain the permutation entropy per symbol of order 6, that is, $h_6 = S/5$ as defined in Bandt and Pompe's work~\cite{bandt2002permutation}. Figure~\ref{fig:3}{a} depicts the well-known bifurcation diagram for the logistic map, while Fig.~\ref{fig:3}{b} shows the values of $h_6$ as a function of the parameter $r$. We note that the permutation entropy per symbol has an overall increasing trend with the parameter $r$, marked by abrupt drops in intervals of $r$ related to periodic behaviors. As noticed by Bandt and Pompe, the behavior of the permutation entropy is similar to the one observed for the Lyapunov exponent~\cite{bandt2002permutation}.

In another example with \permutationentropy{}, we replicate a numerical experiment of Cao \textit{et al.}~\cite{cao2004detecting} (see their Fig.~1) that searches for dynamical changes in the transient logistic map time series (see Appendix~\ref{appendix:chaos} for definitions). This problem illustrates the role of the embedding delay $\tau_x$. As in the original article, we iterate the transient logistic map starting with the initial condition $x_0 = 0.65$ and incrementing the logistic parameter $r$ from $2.8$ to $4$ in steps of size $10^{-5}$. This process generates a time series with $120001$ observations as shown in Fig.~\ref{fig:3}{c}. Using this time series, we calculate the normalized permutation entropy within a sliding window with 1024 observations for the embedding dimension $d_x = 5$ and two values for the embedding delay ($\tau_x = 1$ and $\tau_x = 2$). 

As Cao \textit{et al.}~\cite{cao2004detecting}, we denote the permutation entropy values by $H[r(t)]$, where $r(t)$ represents the logistic parameter at the end of the sliding window. Figure~\ref{fig:3}{d} shows the values of $H[r(t)]$, where abrupt changes are clearly associated with dynamical changes observed in the time series (Fig.~\ref{fig:3}{c}). Despite the overall similarities, we note that the embedding delay $\tau_x = 2$ identifies these dynamical changes better than the case with $\tau_x = 1$; for instance, the transition from period-8 to period-16 (at $r \approx 3.56$) is missed when $\tau_x = 1$ but captured when $\tau_x = 2$~\cite{cao2004detecting}.

As we have mentioned, a generalization of permutation entropy to two-dimensional data was first proposed by Ribeiro \textit{et al.}~\cite{ribeiro2012complexity}. To illustrate the use of the \permutationentropy{} function with two-dimensional data, we replicate a numerical experiment related to Ising surfaces (see Appendix~\ref{appendix:stochastic} for definitions) present in that work (Fig.~8 of Ref.~\onlinecite{ribeiro2012complexity}). These surfaces represent the accumulated sum of spin variables of the canonical two-dimensional Ising model in a Monte Carlo simulation. Figure~\ref{fig:3}{e} shows four examples of these surfaces (square lattices of size $250 \times 250$) obtained after $10^6$ Monte Carlo steps for different reduced temperatures $T_r$. We notice non-trivial patterns emerging when the reduced temperature is equal to the critical temperature $(T_r = 1)$ of phase transition for the Ising model~\cite{landau2015guide}. Following the original article, we generate Ising surfaces (size $250 \times 250$) for reduced temperatures $T_r \in \{0.5,0.6,\dots,3.0\}$ and calculate their normalized permutation entropy with $d_x = 3$ and $d_y = 2$, and $d_x = 2$ and $d_y = 3$, both for $\tau_x = \tau_ y = 1$. In agreement with Ribeiro \textit{et al.}~\cite{ribeiro2012complexity}, Fig.~\ref{fig:3}{f} shows that the permutation entropy precisely identifies the phase transition of the Ising model (the sudden decrease around the critical temperature) and that these Ising surfaces are symmetric under reversal of the embedding dimensions.

\begin{figure*}[!ht]
\centering
\includegraphics[width=0.8\linewidth]{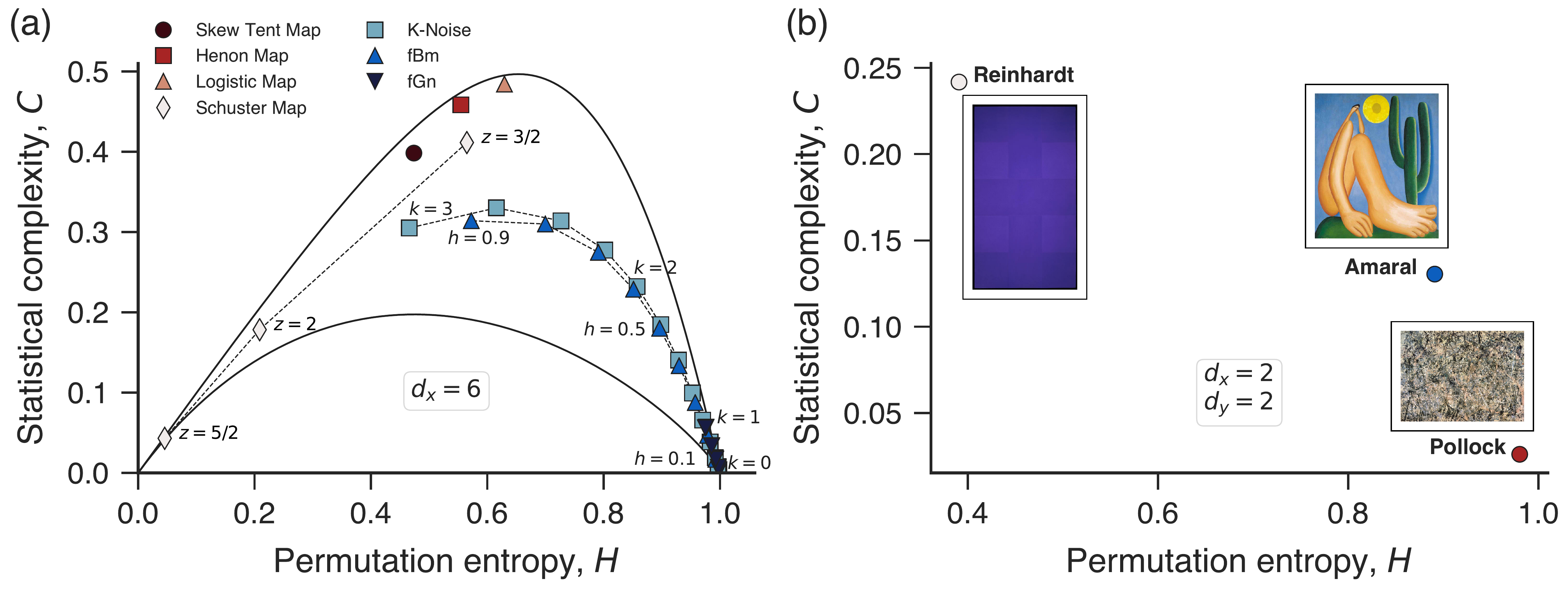}
\caption{Complexity-entropy plane for one- and two-dimensional data. (a) Average values of the statistical complexity $C$ versus the normalized permutation entropy $H$ (over ten realizations) evaluated from time series of chaotic maps and stochastic processes. The embedding parameters are $d_x = 6$ and $\tau_x = 1$. The solid  lines represent the maximum and minimal possible values of complexity for a given entropy (for $d_x = 6$ and $\tau_x = 1$). (b) Localization of three art paintings in the complexity-entropy plane with embedding parameters $d_x = d_y = 2$ and $\tau_x = \tau_y = 1$. All data and code necessary to reproduce this figure are available in a Jupyter notebook at \href{http://github.com/arthurpessa/ordpy}{\ordpy{}}'s webpage.}
\label{fig:4}
\end{figure*}

The \complexityentropy{} function simultaneously calculates the permutation entropy and the statistical complexity from time series and image data. To illustrate its usage, we partially reproduce the results of Rosso \textit{et al.}~\cite{rosso2007distinguishing} (Fig.~1 in that work) on distinguishing chaotic from stochastic time series. By following their article, we iterate four discrete maps to generate chaotic series. Specifically, we obtain chaotic time series from skew tent map (parameter $w = 0.1847$), H\'enon map ($x$-component, parameters $a = 1.4$ and $b = 0.3$), logistic map $(r = 4)$, and Schuster map (parameter $z \in \{3/2, 2, 5/2\}$) -- see Appendix~\ref{appendix:chaos} for definitions. We further generate stochastic series from three stochastic processes: noises with $1/f^{-k}$ power spectrum (for $k \in \{0.00, 0.25, \dots, 3.00\}$), fractional Brownian motion (Hurst exponent $h \in \{0.1,0.2,\dots,0.9\}$), and fractional Gaussian noise (also $h \in \{0.1,0.2,\dots,0.9\}$) -- see Appendix~\ref{appendix:stochastic} for definitions. For each of these maps and stochastic processes, we generate ten time series with $2^{15}$ observations and random initial conditions. Next, we use \complexityentropy{} with embedding parameters $d_x = 6$ and $\tau_x = 1$ to calculate their statistical complexity and permutation entropy (average values over 10 time series realizations).

As in the original work of Rosso \textit{et al.}~\cite{rosso2007distinguishing}, Fig.~\ref{fig:4}{a} shows that chaotic series usually have high complexity and low entropy values. Stochastic time series, in turn, display high entropy and intermediary complexity values. It is also interesting to note that stochastic time series approach the lower-right corner of the complexity-entropy plane ($H\to1$ and $C\to0$) as the serial auto-correlation decreases~\cite{rosso2007distinguishing}. These results also illustrate that some stochastic and chaotic series have very similar entropy values but different statistical complexity (for instance, fractional Brownian motion with $h=0.9$ and Schuster map with $z=3/2$), confirming that the statistical complexity extracts additional information from the ordinal distribution. In this figure, we have also included two solid lines delimiting the accessible region of the complexity-entropy plane~\cite{martin2006generalized}. In \ordpy{}, the functions \maxcurveentropy{} and \mincurveentropy{} generate these curves, as shown in the following code snippet:
\begin{minted}{python}
>>> from ordpy import 
... maximum_complexity_entropy, 
... minimum_complexity_entropy
>>> maximum_complexity_entropy(dx=4)
array([[-0.        , -0.        ],
       [ 0.21810429,  0.19670592],
       [ 0.34568712,  0.28362016],
       ...
       [ 0.98660828,  0.02388382]])
>>> minimum_complexity_entropy(dx=4)
array([[-0.00000000e+00, -0.00000000e+00],
       [ 2.67076969e-02,  2.55212327e-02],
       ...
       [ 1.00000000e+00, -3.66606083e-16]])
\end{minted}

The \complexityentropy{} function also works with two-dimensional data, and to illustrate its usage, we follow Sigaki \textit{et al.}~\cite{sigaki2018history} and use the complexity-entropy plane to investigate patterns in art paintings. Due to the large-scale of the data analyzed by Sigaki \textit{et al.} and to keep our examples self-contained, we do not reproduce their original results but simply use their ideas to illustrate how complexity and entropy extract useful information from images. To do so, we handpick three paintings from \href{https://www.wikiart.org/}{wikiart.org} (in the original article, the authors studied 137,364 images obtained from the same webpage). These are a Color Field Painting artwork (Blue, 1953 by Ad Reinhardt, image size $768 \times 435$~\cite{reinhardt1953abstract}), a Brazilian Modernist artwork (Abaporu, 1928 by Tarsila do Amaral, image size $1200 \times 1026$~\cite{tarsila1928abaporu}), and an American Abstract Expressionist painting (Number 1, 1950 (Lavender Mist), 1950 by Jackson Pollock, image size $749 \times 1024$~\cite{pollock1950number}). The three images are in JPEG format with 24 bits per pixel (8 bits for red, green, and blue colors in the RGB color space). We have averaged the pixels over the three RGB layers to represent each image by a usual two-dimensional array. Having these arrays, we calculate the statistical complexity and permutation entropy for the three paintings with embedding parameters $d_x = d_y = 2$ and $\tau_x = \tau_y = 1$.

Figure~\ref{fig:4}{b} shows the complexity-entropy plane for these images (insets depict the artworks). In agreement with the global trend observed by Sigaki \textit{et al.}~\cite{sigaki2018history}, these results show that paintings portraying objects with clearly defined borders (such as the squares in Reinhardt's artwork) tend to present large values of statistical complexity and low values of entropy. On the other extreme, paintings with smudged and diffuse contours (such as Pollock's drip paintings) have high entropy and low complexity values. Between these somewhat opposite behaviors, we have a whole continuum of images, as exemplified here by the work of the Brazilian painter Tarsila do Amaral. As argued by Sigaki \textit{et al.}~\cite{sigaki2018history}, the complexity-entropy plane maps the local degree of order of artworks into a scale of order-disorder and simplicity-complexity that is similar to qualitative descriptions of artworks proposed by art historians such as W\"olfflin (the linear versus painterly dichotomy) and Riegl (the haptic versus optic dichotomy).

\section{Missing ordinal patterns}

As we have commented, the logistic map at fully developed chaos does not exhibit the ``descending permutation'' $(2,1,0)$ for $d=3$ (see Fig.~\ref{fig:2}{b}). This feature is not a particularity of the logistic map. Indeed, these missing ordinal patterns (also called forbidden patterns) occur in different systems, and simple statistics associated with them have proven to be useful and reliable indicators of a system's dynamics~\cite{zanin2008forbidden, zunino2009forbidden, sakellariou2016counting2, mcculough2016counting1, kulp2016using}. The works of Amig\'o \textit{et al.}~\cite{amigo2006order, amigo2007true} are seminal in this regard, and by following their classification, we can divide these forbidden ordinal patterns into two categories: true or false~\cite{amigo2007true}. True forbidden patterns (such as the  $(2,1,0)$ in the logistic map) are a fingerprint of determinism in a time series dynamics and represent an intrinsic feature of the underlying dynamical process~\cite{amigo2006order}; that is, these patterns are not an artifact related to the finite length of empirical observations. In turn, false forbidden patterns are related to the finite length of time series~\cite{amigo2007true} and can emerge even from completely random processes.

\begin{figure*}[!t]
\centering
\includegraphics[width=0.8\linewidth]{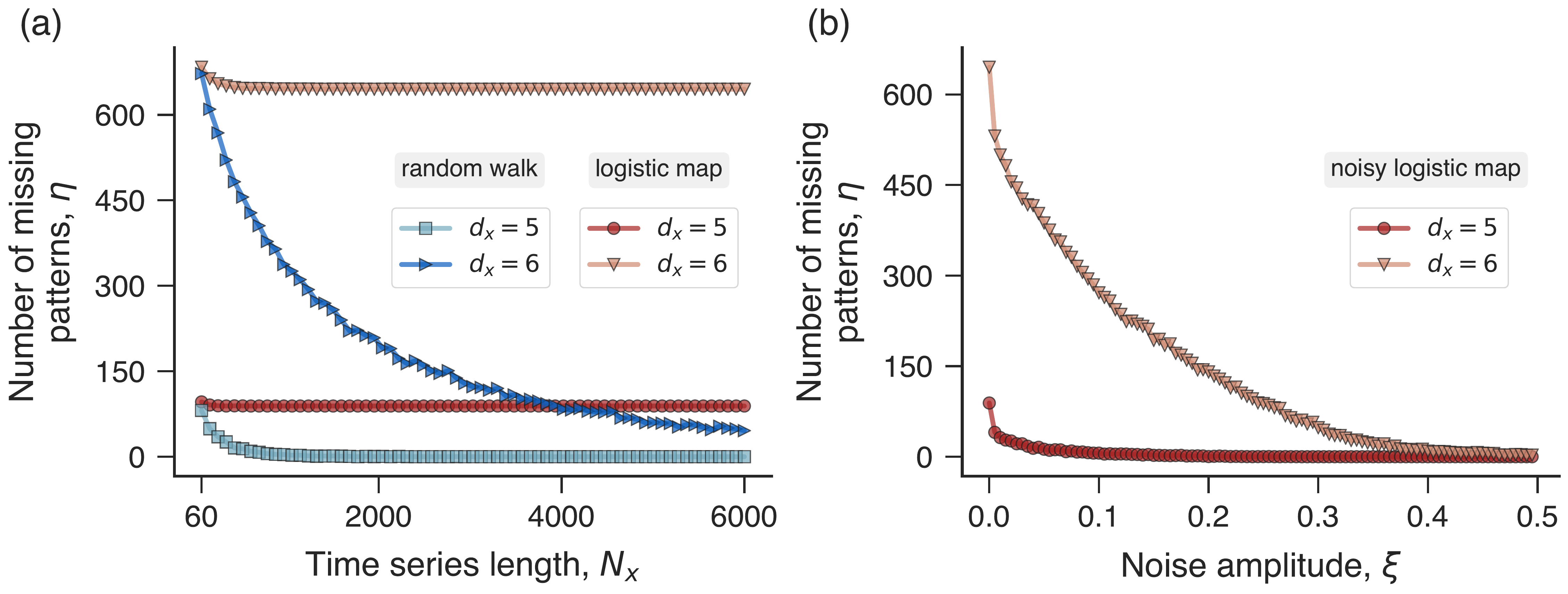}
\caption{Missing ordinal patterns in time series. (a) Number of missing ordinal patterns ($\eta$) in random walk (blue) and logistic map (red) time series as a function of sequence length ($N_x$) for embedding parameters $d_x = 5$ and $d_x = 6$, both with $\tau_x = 1$. Results represent the average number of missing permutations over ten time series replicas for each $N_x\in \{60,150,240,\dots,6000\}$. (b) Dependence of the number of missing ordinal patterns on the noise intensity ($\xi$) for noisy logistic time series with 6000 observations. The noise added to the logistic series is uniformly distributed in the interval $[-\xi, \xi]$ with $\xi \in \{0,0.001,0.002,\dots,0.5\}$. Results represent average values over ten time series replicas for each noise level. The embedding dimensions are indicated within the plot and the embedding delay is $\tau_x = 1$. We use random initial conditions and set the parameter $r = 4$ in all experiments with the logistic map. The necessary code to reproduce these results is available in a Jupyter notebook at \href{http://github.com/arthurpessa/ordpy}{\ordpy{}}'s webpage.}
\label{fig:5}
\end{figure*}

This distinction is not straightforward when dealing with empirical data, but a typical analysis in this context consists in investigating the number of missing patterns ($\eta$) as a function of the time series length ($N_x$). The behavior of this curve is useful for discriminating time series. In \ordpy{}, the \missingstates{} function identifies missing ordinal patterns and estimates their relative frequency as in:
\begin{minted}{python}
>>> from ordpy import missing_patterns
>>> missing_patterns([4,7,9,10,6,11,3,5,
... 6,2,3,1], dx=3)
(array([[0, 2, 1],
        [2, 1, 0]]),
 0.3333333333333333)
\end{minted}

To better illustrate the use of this function, we investigate missing ordinal patterns in time series obtained from the logistic map at fully developed chaos ($r = 4$) and Gaussian random walks. In both cases, we use the embedding dimensions $d_x = 5$ and $d_x = 6$ (with $\tau_x = 1$) and series lengths $N_x \in \{60,150,240,\dots,6000\}$. Figure~\ref{fig:5}{a} shows the results. We observe that the number of missing ordinal patterns approaches zero as the time series length of random walks increases. Conversely, the number of missing permutations related to the logistic map displays an initial decay with the time series length but it rapidly saturates in considerably large numbers, indicating that these missing patterns are intrinsically associated with the underlying determinism of the process~\cite{amigo2007true}.

In another application with the \missingstates{} function, we replicate a result of Amig\'o \textit{et al.}~\cite{amigo2007true} (Fig.~4 in their work) to further show that the number of missing patterns is a good indicator of determinism in time series~\cite{amigo2007true, amigo2008combinatorial}. By following the original work, we generate time series from the logistic map at fully developed chaos (6000 iterations) and add to them uniformly distributed noise in the interval $[-\xi, \xi]$, where $\xi$ is the noise amplitude. Next, we estimate the average number of missing patterns (over ten time series replicas) for each noise level $\xi \in \{0,0.001,0.002,\dots,0.5\}$, and embedding dimensions $d_x=5$ and $d_x=6$ (with $\tau_x=1$). Figure~\ref{fig:5}{b} shows the number of missing ordinal patterns as a function of noise amplitude $\xi$ for both embedding dimensions. We observe that the number of missing patterns related to these deterministic series contaminated with noise approaches zero as noise amplitude grows. However, significantly higher noise levels are necessary to remove all signs of determinism expressed by the lack of permutation patterns when $d_x=6$~\cite{amigo2007true}.

\section{Tsallis and R\'enyi entropy-based quantifiers of the ordinal distribution}

In addition to Shannon's entropy and the statistical complexity, researchers have proposed to use other quantifiers of the ordinal probability distribution~\cite{bandt2017new, small2018ordinal, zunino2008fractional, liang2015eeg}. As we have explicitly verified for the statistical complexity, these different quantifiers are supposed to extract additional information from a time series' dynamics that is not captured by permutation entropy and statistical complexity. In this context, a productive approach is to consider parametric generalizations of Shannon's entropy, such as those proposed by Tsallis~\cite{tsallis1988generalization} and R\'enyi~\cite{renyi1961measures}. The work of Zunino \textit{et al.}~\cite{zunino2008fractional} was the first to consider the Tsallis entropy in place of Shannon's entropy to define the Tsallis permutation entropy as
\begin{equation}
    S_\beta(P) = \frac{1}{\beta-1}\sum_{i = 1}^{n_\pi} (\rho_i(\Pi_i) - \rho_i(\Pi_i)^\beta)\,,
\end{equation}
where $\beta$ is a real parameter ($\beta \to 1$ recovers the usual Shannon entropy and so the permutation entropy). Tsallis's entropy is also maximized by the uniform distribution, such that $S_\beta^{\rm max} = \frac{1 - (n_\pi)^{1-\beta}}{\beta-1}$. Thus, the normalized Tsallis permutation entropy is
\begin{equation}
    H_\beta(P) = (\beta-1)\frac{S_\beta(P)}{1 - (n_\pi)^{1-\beta}}.
\end{equation}

Similarly, Liang \textit{et al.}~\cite{liang2015eeg} have proposed the R\'enyi permutation entropy
\begin{equation}
    S_\alpha(P) = \frac{1}{1-\alpha}\ln\left(\sum_{i = 1}^{n_\pi} \rho_i(\Pi_i)^\alpha\right)\,,
\end{equation}
where $\alpha > 0$ is a real parameter. R\'enyi's entropy converges to Shannon's entropy when $\alpha \to 1$ and is maximized by the uniform distribution ($S_\alpha^{\rm max} = \ln n_\pi$, as the usual Shannon entropy). Thus, the normalized R\'enyi permutation entropy is
\begin{equation}
    H_\alpha(P) = \frac{S_\alpha(P)}{\ln n_\pi}.
\end{equation}

In both cases, the generalized entropic form is mono-parametric and has a term where the ordinal probabilities appear raised to the power of the entropic parameter (that is, $\rho_i(\Pi_i)^\beta$ and $\rho_i(\Pi_i)^\alpha$). These parameters assign different weights to the underlying ordinal probabilities, allowing us to access different dynamical scales and produce a family of quantifiers for the ordinal distribution. In \ordpy{}, the \tsallisentropy{} and \renyientropy{} functions implement these two quantifiers as in:
\begin{minted}{python}
>>> from ordpy import tsallis_entropy, 
... renyi_entropy 
>>> tsallis_entropy([4,7,9,10,6,11,3], 
... q=[1,2], dx=2) #Here q plays the
... role of beta.
array([0.91829583, 0.88888889])
>>> renyi_entropy([4,7,9,10,6,11,3],
... alpha=[1,2], dx=2)
array([0.91829583, 0.84799691])
\end{minted}

In a similar direction, there are also the developments of complexity-entropy curves proposed by Ribeiro \textit{et al.}~\cite{ribeiro2017characterizing} and Jauregui \textit{et al.}~\cite{jauregui2018characterization}. These works have further extended the complexity-entropy plane concept by considering the Tsallis and R\'enyi entropies combined with proper generalizations of statistical complexity~\cite{martin2006generalized}. Thus, instead of having a single point in the complexity-entropy plane for a given time series, Ribeiro \textit{et al.}~\cite{ribeiro2017characterizing} and Jauregui \textit{et al.}~\cite{jauregui2018characterization} have created parametric curves by varying the entropic parameter ($\beta$ or $\alpha$) and simultaneously calculating the generalized entropy and the generalized statistical complexity. 

To define the Tsallis complexity-entropy curves~\cite{ribeiro2017characterizing}, we first extend the statistical complexity (Eq.~\ref{eq:statistical_complexity}) using the Tsallis entropy, that is,
\begin{equation}\label{eq:statistical_complexity_tsallis}
    C_\beta(P) = \frac{D_\beta(P,U) H_\beta(P)}{D^{\rm max}_\beta}\,,
\end{equation}
where
\begin{equation}
    D_\beta(P,U) = \frac{1}{2}K_\beta\left( P\bigg|\frac{P+U}{2}\right) + \frac{1}{2}K_\beta\left(U\bigg|\frac{P+U}{2}\right)
\end{equation}
is the Jensen-Tsallis divergence~\cite{martin2006generalized} written in terms of the corresponding Kullback-Leibler divergence~\cite{martin2006generalized, tsallis2009introduction}
\begin{equation}
    K_\beta(V|R) = \frac{1}{\beta-1}\sum_i^{n_\pi} v_i^\beta[r_i^{1-\beta}-v_i^{1-\beta}]\,,
\end{equation}
where $V=\{v_i\}_{i = 1, \dots, n_\pi}$ and $R=\{r_i\}_{i = 1, \dots, n_\pi}$ are two arbitrary distributions. In Eq.~\ref{eq:statistical_complexity_tsallis}, \begin{equation*}
    D^{\rm max}_\beta = \frac{2^{2-\beta}n_\pi - (1 + n_\pi)^{1-\beta}-n_\pi(1+1/n_\pi)^{1-\beta}-n_\pi+1}{2^{2-\beta}n_\pi(1-\beta)}
\end{equation*}
is a normalization constant representing the maximum possible value of $D_\beta(P,U)$ that occurs for $P = \{ \delta_{1,i} \}_{i = 1,\dots,n_\pi}$ (as in the usual Jensen-Shannon divergence). By following Ribeiro \textit{et al.}~\cite{ribeiro2017characterizing}, we construct a parametric representation of the ordered pairs $(H_\beta(P), C_\beta(P))$ for $\beta > 0$, obtaining the Tsallis complexity-entropy curves. 

Similarly, to define the R\'enyi complexity-entropy curves~\cite{jauregui2018characterization}, we generalize the statistical complexity in R\'enyi's formalism as 
\begin{equation}
    C_\alpha(P,U) = \frac{D_\alpha(P,U) H_\alpha(P)}{D_\alpha^{\rm max}}\,,
\end{equation}
where
\begin{equation}
    D_\alpha(P,U) = \frac{1}{2}K_\alpha\left( P\bigg|\frac{P+U}{2}\right) + \frac{1}{2}K_\alpha\left(U\bigg|\frac{P+U}{2}\right)
\end{equation}
is the Jensen-R\'enyi divergence~\cite{martin2006generalized} written in terms of 
\begin{equation}
    K_\alpha(V|R) = \frac{1}{\alpha-1}\ln \left( \sum_{i = 1}^{d!} v_i^\alpha r_i^{1-\alpha}\right)\,,    
\end{equation}
the corresponding Kullback-Leibler divergence for R\'enyi's entropy~\cite{martin2006generalized, vanerven2014renyi}. The normalization constant
\begin{equation*}
    D_\alpha^{\rm max} = \frac{1}{2(\alpha-1)}\ln{\left[\frac{(n_\pi+1)^{1-\alpha} + n_\pi - 1}{n_\pi}\left(\frac{n_\pi+1}{4 n_\pi}\right)^{1-\alpha}\right]}
\end{equation*}
corresponds to the maximum possible value of $D_\alpha(P,U)$ occurring for $P = \{ \delta_{1,i} \}_{i = 1,\dots,n_\pi}$ (as in the usual Jensen-Shannon divergence). Again, we can construct a parametric representation of the ordered pairs $(H_\alpha(P), C_\alpha(P))$ for $\alpha > 0$, obtaining the R\'enyi complexity-entropy curves proposed by Jauregui \textit{et al.}~\cite{jauregui2018characterization}.

In \ordpy{}, the functions \tsallis{} and \renyi{} implement the Tsallis and R\'enyi complexity-entropy curves as shown in the following code snippet:
\begin{minted}{python}
>>> from ordpy import 
... tsallis_complexity_entropy,
... renyi_complexity_entropy
>>> tsallis_complexity_entropy(
... [4,7,9,10,6,11,3], 
... dx=2, q=[1,2]) #Here q plays the
... role of beta.
array([[0.91829583, 0.06112817],
       [0.88888889, 0.07619048]])
>>> renyi_complexity_entropy(
... [4,7,9,10,6,11,3],
... dx=2, alpha=[1, 2])
array([[0.91829583, 0.06112817],
       [0.84799691, 0.08303895]])
\end{minted}

To better illustrate the use of these \ordpy{}'s functions, we replicate some numerical experiments involving the logistic map and random walks presented in the original works of Ribeiro \textit{et al.}~\cite{ribeiro2017characterizing} (Figs.~1 and 6 in that work) and Jauregui \textit{et al.}~\cite{jauregui2018characterization} (Figs.~1 and 3 in that work). We start by generating time series from the logistic map at fully developed chaos ($r=4$, random initial condition) and a Gaussian random walk. For the logistic map series, we discard the first $10^4$ iterations to avoid transient effects and iterate other $10^6$ steps. The random walk series also has $10^6$ observations. By using these time series, we generate their corresponding Tsallis complexity-entropy curves for  $d_x = 3$ and $\tau_x = 1$ by sampling $10^3$ log-spaced values of the entropic parameter $\beta$ between $0.01$ and $100$ for the logistic map, and between $0.001$ and $100$ for the random walk. 

\begin{figure*}[!t]
\centering
\includegraphics[width=0.8\linewidth]{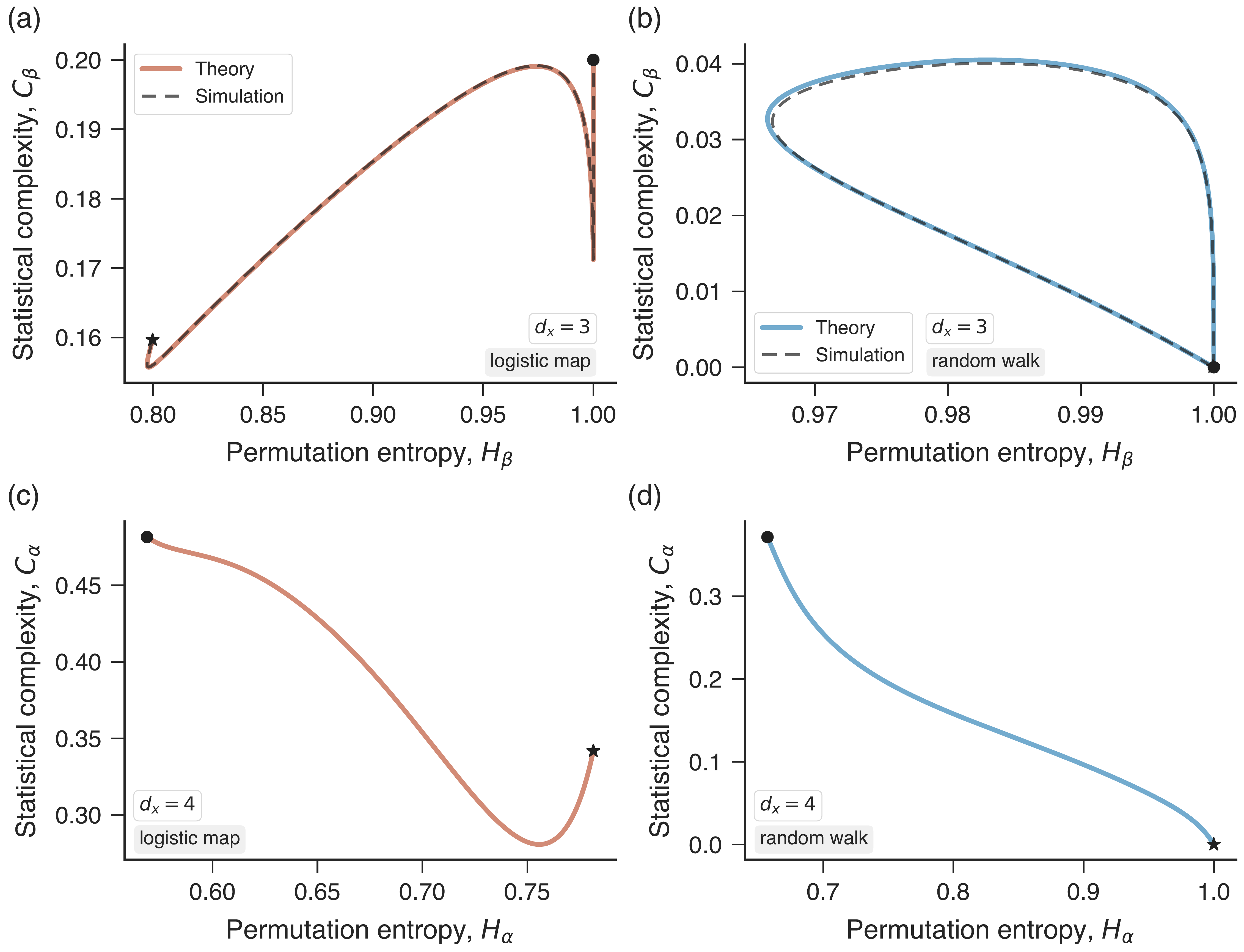}
\caption{Tsallis and R\'enyi complexity-entropy curves. Tsallis complexity-entropy curves for time series obtained from (a) the logistic map at fully developed chaos and (b) a Gaussian random walk, both with embedding parameters $d_x = 3$ and $\tau_x = 1$. The solid lines represent the empirical results and the dashed lines indicate the exact form of these complexity-entropy curves. Panels (c) and (d) show the R\'enyi complexity-entropy curves obtained from the same two time series with embedding parameters $d_x = 4$ and $\tau_x = 1$. In all panels, star markers indicate the beginning of the curves ($\beta\approx0$ or $\alpha\approx0$), while circle markers indicate the end of the curves (largest values of $\beta$ and $\alpha$). Data and code necessary to reproduce these results are available in a Jupyter notebook at \href{http://github.com/arthurpessa/ordpy}{\ordpy{}}'s webpage.}
\label{fig:6}
\end{figure*}

Figures~\ref{fig:6}{a} and \ref{fig:6}{b} show the empirical complexity-entropy curves in comparison with their exact shape (dashed lines). These theoretical curves can be determined for these time series because the ordinal distributions of the logistic map ($P_{\rm logistic} = \{{1}/{3}, {1}/{15}, {2}/{15}, {3}/{15}, {4}/{15}, 0\}$) and random walks ($P_{\rm walk} = \{{1}/{4}, {1}/{8}, {1}/{8}, {1}/{8}, {1}/{8}, {1}/{4}\}$) are exactly known for $d_x = 3$~\cite{amigo2006order, bandt2007order}. We observe that theoretical and empirical results are in excellent agreement. As discussed by Ribeiro \textit{et al.}~\cite{ribeiro2017characterizing}, random series tend to form closed complexity-entropy curves (Fig.~\ref{fig:6}{b}), while chaotic time series are usually represented by open complexity-entropy curves (Fig.~\ref{fig:6}{a}). These features emerge as a direct consequence of the existence or not of missing ordinal patterns captured by the limiting behavior of $H_\beta$ as $\beta \to 0$ and $\beta \to \infty$~\cite{ribeiro2017characterizing}.

By following a similar approach, we also estimate the R\'enyi complexity-entropy curves for the two previous time series for $d_x = 4$ and $\tau_x = 1$. Figures~\ref{fig:6}{c} and \ref{fig:6}{d} show these R\'enyi complexity-entropy curves. Differently from the Tsallis case, R\'enyi complexity-entropy curves are always open~\cite{jauregui2018characterization}, and the usage of these curves for distinguishing chaotic from stochastic series relies on a more subtle characteristic. Indeed, Jauregui \textit{et al.}~\cite{jauregui2018characterization} have found that the initial curvature of R\'enyi complexity-entropy curves ($dC_\alpha/dH_\alpha$ for small $\alpha$) can be used as an indicative of determinism in time series. Specifically, they found that positive curvatures are associated with time series of stochastic nature, while negative ones are related to chaotic phenomena. This pattern also occurs in the results of Figs.~\ref{fig:6}{c} and \ref{fig:6}{d}.

\section{Ordinal networks}

Among the more recent developments related to the Bandt-Pompe framework, we have the so-called ordinal networks. First proposed by Small~\cite{small2013complex} for investigating nonlinear dynamical systems, and later generalized with his collaboration in a series of works~\cite{mccullough2015time, mccullough2017multiscale, sun2014characterizing, sakellariou2019markov}, ordinal networks belong to a more general class of methods designed to map time series into networks, collectively known as time series networks~\cite{zou2019complex}. Beyond counting ordinal patterns, this approach considers first-order transitions among ordinal symbols within a symbolic sequence. In this network representation, the different ordinal patterns occurring in a data set are mapped into nodes of a complex network. The edges between nodes indicate that the associated permutation symbols are adjacent to each other in a symbolic sequence. Furthermore, edges can be directed according to the temporal succession of ordinal symbols and weighted by the relative frequencies in which the corresponding successions occur in a symbolic sequence~\cite{mccullough2015time}.

After applying the Bandt-Pompe method with embedding parameters $d_x$ and $\tau_x$ to a time series $\{x_t\}_{t=1,\dots,N_x}$ and obtaining the symbolic sequence $\{\pi_p\}_{p = 1, \dots, n_x}$, we can define the elements of the weighted adjacency matrix of the corresponding ordinal network as~\cite{mccullough2015time, pessa2019characterizing}
\begin{equation}\label{eq:adj_matrix}
    \rho_{i, j} = \frac{\text{total of transitions} \ \Pi_i \to \Pi_j\ \text{in}\ \{\pi_p\}_{p = 1, \dots, n_x}}{n_x - 1}\,,
\end{equation} 
where $i,j=1,2,\dots,n_\pi$ (with $n_\pi=d_x!$), $\Pi_i$ and $\Pi_j$ represent all possible ordinal patterns, and the denominator $n_x - 1$ is the total number of ordinal transitions. In \ordpy{}, the \ordinalnetwork{} function returns the nodes, edges, and edge weights of an ordinal network mapped from a time series as in:
\begin{minted}{python}
>>> from ordpy import ordinal_network
>>> ordinal_network([4,7,9,10,6,11,8,3,7],
... dx=2, normalized=False)
(array(['0|1', '1|0'], dtype='<U3'),
 array([['0|1', '0|1'],
        ['0|1', '1|0'],
        ['1|0', '0|1'],
        ['1|0', '1|0']], dtype='<U3'),
 array([2, 2, 2, 1]))
\end{minted}

It is worth noting that the original algorithm of Small~\cite{small2013complex} for mapping time series into ordinal networks uses a different approach for creating the symbolic sequence. Instead of defining overlapping partitions (Eq.~\ref{eq:1dpartition}), Small~\cite{small2013complex} evaluates the ordinal patterns in non-overlapping partitions of size $d_x$ (the embedding delay is also not present in his original formulation). Furthermore, edges are undirected and unweighted in this initial formulation. This implementation is not as popular as the one directly following the Bandt-Pompe symbolization method~\cite{mccullough2015time, mccullough2017multiscale, sun2014characterizing, sakellariou2019markov}, but is also available in \ordpy{} through the \texttt{overlapping} parameter in the \ordinalnetwork{} function, as shown in:
\begin{minted}{python}
>>> from ordpy import ordinal_network
>>> ordinal_network([4,7,9,10,6,11,8,3,7],
... dx=2, normalized=False,
... overlapping=False)
(array(['0|1', '1|0'], dtype='<U3'), 
array([['0|1', '0|1'], ['0|1', '1|0']], 
dtype='<U3'), array([2, 1]))
\end{minted}

Ordinal networks have also been recently generalized by Pessa and Ribeiro~\cite{pessa2020mapping} to account for two-dimensional data sets such as images. In this case, we apply the two-dimensional version of Bandt and Pompe's symbolization approach~\cite{ribeiro2012complexity} (see Eqs.~\ref{eq:matrix_partition}, \ref{eq:partition_index}, and \ref{eq:partition}) to a data array $\{y_t^u\}_{t = 1,\dots,N_x}^{u = 1,\dots,N_y}$ for given embedding dimensions ($d_x$ and $d_y$) and embedding delays ($\tau_x$ and $\tau_y$), obtaining the corresponding two-dimensional ordinal sequence $\{\pi_p^q\}_{p = 1,\dots,n_x}^{q = 1,\dots,n_y}$. Similarly to the one-dimensional case, each permutation symbol $\Pi_i$ ($i = 1,\dots,n_\pi$, with $n_\pi=(d_x d_y)!$) is associated with a node in the ordinal network, and directed edges connect permutation symbols that are vertically ($\pi_p^q \to \pi_p^{q+1}$ for $q = 1,\dots,n_y-1$) or horizontally ($\pi_p^q \to \pi_{p+1}^{q}$ for $p = 1,\dots,n_x-1$) adjacents in the symbolic sequence. The directed link between a pair of permutation symbols ($\Pi_i$ and $\Pi_j$) is weighted by the total number of occurrences of this particular transition in the symbolic sequence. Thus, the weighted adjacency matrix representing the ordinal network mapped from two-dimensional data is~\cite{pessa2020mapping}
\begin{equation}\label{eq:edge_weights}
    \rho_{i,j}\! =\! \frac{\text{total of transitions $\Pi_i \to \Pi_j$ in $\{\pi_p^q\}_{p=1,\dots,n_x}^{q=1,\dots,n_y}$}}{2n_x n_y-n_x-n_y}\,,
\end{equation}
where $i, j = 1,\dots,n_\pi$ (with $n_\pi = (d_x d_y)!$) and the denominator represents the total number of horizontal and vertical transitions. The \ordinalnetwork{} function also handles two-dimensional data as in:
\begin{minted}{python}
>>> from ordpy import ordinal_network
>>> ordinal_network([[1,2,1],[8,3,4],[6,7,5]], 
... dx=2, dy=2, normalized=False)
(array(['0|1|3|2', '1|0|2|3', '1|2|3|0'], 
dtype='<U7'),
 array([['0|1|3|2', '1|0|2|3'],
        ['0|1|3|2', '1|2|3|0'],
        ['1|0|2|3', '0|1|3|2'],
        ['1|2|3|0', '0|1|3|2']], dtype='<U7'),
 array([1, 1, 1, 1]))
\end{minted}

Pessa and Ribeiro~\cite{pessa2020mapping} have also proposed to create ordinal networks by considering only horizontal (horizontal ordinal networks) or only vertical (vertical ordinal networks) transitions among the permutations symbols. They have shown that comparing properties of these two networks is useful for exploring visual symmetries in images. In \ordpy{}, this possibility is available through the \texttt{connections} parameter in the \ordinalnetwork{} function as in: 
\begin{minted}{python}
>>> from ordpy import ordinal_network
>>> ordinal_network([[1,2,1],[8,3,4],[6,7,5]],
... dx=2, dy=2, normalized=False, 
... connections='horizontal')
(array(['0|1|3|2', '1|0|2|3', '1|2|3|0'], 
dtype='<U7'), 
array([['0|1|3|2', '1|0|2|3'],
       ['1|2|3|0', '0|1|3|2']], dtype='<U7'), 
array([1, 1]))
\end{minted}

An intriguing feature of ordinal networks is the existence of intrinsic connectivity constraints~\cite{pessa2019characterizing, pessa2020mapping} inherited from Bandt and Pompe's symbolization method. These constraints are directly related to the fact that adjacent partitions share elements, such that ordering relations in one partition are partially carried out to neighboring partitions. For one-dimensional data, these restrictions imply that all nodes in an ordinal network have in-degree and out-degree limited to numbers between $0$ and $d_x$; consequently, the maximum number of edges is $d_x \times (d_x!)$~\cite{pessa2019characterizing}. Ordinal networks mapped from one-dimensional data can only have self-loops in nodes associated with solely ascending or solely descending ordinal patterns~\cite{pessa2019characterizing}. 

The horizontal and vertical transitions related to networks mapped from two-dimensional data impose similar but trickier connectivity constraints~\cite{pessa2020mapping}. In this case, the maximum number of outgoing connections emerging from horizontal and vertical transitions are $C(d_x d_y, d_y) \times d_y!$ and $C(d_x d_y, d_x) \times d_x!$, respectively~\cite{pessa2020mapping}. However, the sets of horizontal and vertical transitions are not disjoint, and their union defines all possible outgoing edges. Finding a general expression for the latter set operation is cumbersome because it depends on the ordinal pattern associated with the node under analysis. Thus, while limited, the maximum number of edges varies among the ordinal patterns and needs to be numerically obtained~\cite{pessa2020mapping}. Furthermore, differently from the one-dimensional case, ordinal networks mapped from two-dimensional data can display self-loops in several nodes~\cite{pessa2020mapping}.

A direct consequence of these intrinsic connectivity constraints is that ordinal networks mapped from completely random arrays (in one or two dimensions) are not random graphs~\cite{pessa2019characterizing, pessa2020mapping}. Even more counter-intuitive is the existence of different edge weights in random ordinal networks, albeit all permutations are equiprobable in random arrays~\cite{pessa2019characterizing, pessa2020mapping}. This non-trivial property results from the fact that, among all possible amplitude relations involved in an ordinal transition between a fixed permutation and all its possible neighboring permutations, some permutations appear more than once. For one-dimensional data, random ordinal networks only have two different edge weights: $1/(d_x+1)!$ and $2/(d_x+1)!$ (the denominator represents the sum of weights)~\cite{pessa2019characterizing}. A rule of thumb for determining the edges with double weight is to pick all transitions in which the index number equal to ``$d_x - 1$'' in the next permutation fits the position of the index number ``0'' in the first permutation~\cite{pessa2019characterizing}. For instance, the edge weight between permutations $(3,2,1,0)$ and $(2,1,0,3)$ has double weight. Ordinal networks mapped from two-dimensional random data have more than two different edge weights, and there is no simple rule (at least up to now) for obtaining these weights~\cite{pessa2020mapping}. However, these values can be numerically calculated by explicitly considering each possible ordinal pattern~\cite{pessa2020mapping}.

In \ordpy{}, the \randomordinalnetwork{} function generates the exact form of ordinal networks expected from the mapping of one- and two-dimensional random data with arbitrary embedding dimensions ($d_x$ and $d_y$). The following code illustrates the usage of \randomordinalnetwork{}:
\begin{minted}{python}
>>> from ordpy import random_ordinal_network
>>> random_ordinal_network(dx=2)
(array(['0|1', '1|0'], dtype='<U3'),
 array([['0|1', '0|1'],
        ['0|1', '1|0'],
        ['1|0', '0|1'],
        ['1|0', '1|0']], dtype='<U3'),
 array([0.16666667, 0.33333333, 
        0.33333333, 0.16666667]))
\end{minted}
The three returned arrays represent nodes, edges, and edge weights of the random ordinal network, respectively. It is worth noticing that these connectivity constraints disappear when considering non-overlapping data partitions as in the initial proposal of Small~\cite{small2013complex}. In this case, ordinal networks mapped from large enough random data sets are represented by complete graphs with self-loops and all-equal edge weights. The \randomordinalnetwork{} function returns these graphs by changing its \texttt{overlapping} argument as in: 
\begin{minted}{python}
>>> from ordpy import random_ordinal_network
>>> random_ordinal_network(dx=2,
... overlapping=False)
(array(['0|1', '1|0'], dtype='<U3'),
 array([['0|1', '0|1'],
        ['0|1', '1|0'],
        ['1|0', '0|1'],
        ['1|0', '1|0']], dtype='<U3'),
 array([0.25, 0.25, 0.25, 0.25]))
\end{minted}
Similarly, embedding delays larger than one modify how elements are shared among partitions and impose connectivity constraints to high-order transitions. The \randomordinalnetwork{} function is thus restricted to the case $\tau_x = \tau_y =1$ when considering overlapping partitions.

The primary purpose of mapping time series or images into ordinal networks is to use network measures to characterize data sets. In addition to the many network statistics derived from network science~\cite{newman2010networks}, the inherent probabilistic nature of nodes and edges in ordinal networks has motivated two entropy-related measures~\cite{mccullough2017multiscale, small2018ordinal, pessa2019characterizing}. The first one is a local measure defined at the node level known as the local node entropy~\cite{mccullough2017multiscale, small2018ordinal, pessa2019characterizing, pessa2020mapping}
\begin{equation}\label{eq:local_node_entropy}
    s_i = -\sum_{j\in\mathcal{O}_i}\rho'_{i, j}\log \rho'_{i,j},\,
\end{equation}
where the index $i$ refers to a node related to a given permutation $\Pi_i$, $\rho'_{i,j} = \rho_{i,j}/\sum_{k\in\mathcal{O}_i}\rho_{i,k}$ represents the renormalized probability of transitioning from node $i$ to node $j$ (permutations $\Pi_i$ and $\Pi_j$), and $\mathcal{O}_i$ is the outgoing neighborhood of node $i$ (set of all edges leaving node $i$). This quantity measures the determinism of ordinal transitions at the node level such that $s_i$ is maximum when all edges leaving $i$ have the same weight, while $s_i=0$ if there is only one edge leaving node $i$. Using the local node entropy, we can further define the global node entropy~\cite{mccullough2017multiscale, small2018ordinal, pessa2019characterizing, pessa2020mapping}
\begin{equation}\label{eq:global_node_entropy}
    S_{\rm GN} = \sum_{i=1}^{n_\pi} \rho_i s_i\,,
\end{equation}
where $\rho_i$ is the probability of finding the permutation $\Pi_i$ (Eqs.~\ref{eq:permutation_probability} and \ref{eq:permutation_probability_2d}). Thus, the value $S_{\rm GN}$ represents a weighted average of the local determinism over all nodes of an ordinal network (see also Unakafov and Keller~\cite{unakafov2014conditional} for the definition of conditional entropy of ordinal patterns). In image classification tasks~\cite{pessa2020mapping}, global node entropy has proven to outperform different image quantifiers derived from gray-level co-occurrence matrices (GLCMs)~\cite{haralick1973textural, haralick1979statistical}, a traditional technique for texture analysis.

Contrarily to permutation entropy~\cite{bandt2002permutation} and because of the intrinsic connectivity constraints of ordinal networks, the global node entropy is not maximized by random data~\cite{pessa2019characterizing, pessa2020mapping}. For one-dimensional data, the global node entropy calculated from a random ordinal network is~\cite{pessa2019characterizing}
\begin{equation}
    S_{\rm GN}^{\rm random} = \log(d_x + 1) - (\log 4)/(d_x + 1)\,.
\end{equation}
While there is no equivalent expression for two-dimensional data, it is possible to numerically calculate $S_{\rm GN}^{\rm random}$ using random ordinal networks numerically generated~\cite{pessa2020mapping}. In both cases, the global node entropy can be normalized by the value of $S_{\rm GN}^{\rm random}$, that is, $H_{\rm GN} = {S_{\rm GN}}/{S_{\rm GN}^{\rm random}}$.

\begin{figure*}[!ht]
\centering
\includegraphics[width=1\linewidth]{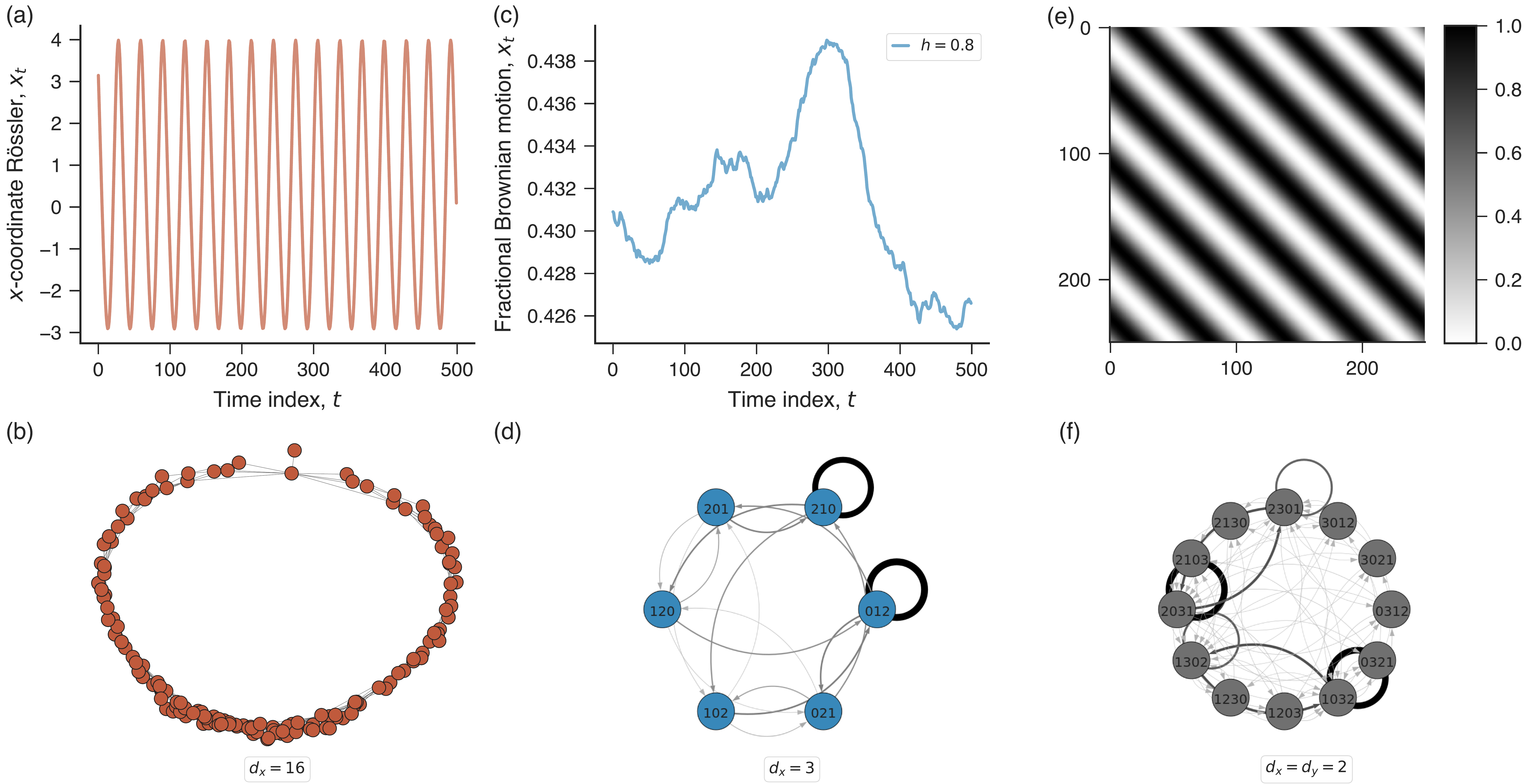}
\caption{Ordinal networks mapped from one- and two-dimensional data. (a) Time series obtained from the $x$-coordinate of the R\"ossler system (with parameters $a=0.3$, $b = 2$ and $c = 4$). Here we show only the latest 500 of all $10^5$ observations. This time series exhibits a periodic behavior after a short initial transient. (b) Visualization of the ordinal network mapped from the $x$-coordinate of the R\"ossler system with embedding parameters $d_x = 16$ and $\tau_x = 1$. This ordinal network uses Small's original algorithm~\cite{small2013complex} with non-overlapping data partitions and undirected and unweighted edges. (c) Time series obtained from a realization of the fractional Brownian motion with $h = 0.8$ (only the latest 500 of all $2^{16}$ observations are shown). (d) Ordinal network representation of the fractional Brownian motion time series with $d_x = 3$ and $\tau_x = 1$. Here we have used overlapping data partitions and made edge thickness proportional to edge weight. (e) Example of a periodic ornament with size $250 \times 250$. (f) Ordinal network representation of the previous image with embedding parameters $d_x = d_y = 2$ and $\tau_x = \tau_y = 1$. In this visualization, edge thickness is made proportional to edge weight. Data and code necessary to reproduce these results are available in a Jupyter notebook at \href{http://github.com/arthurpessa/ordpy}{\ordpy{}}'s webpage.}
\label{fig:7}
\end{figure*}

In \ordpy{}, the \globalnodeentropy{} function evaluates $S_{\rm GN}$ directly from data arrays or using an ordinal network as returned by \ordinalnetwork{}. The following code shows simple usages of \globalnodeentropy{}:
\begin{minted}{python}
>>> from ordpy import global_node_entropy
>>> global_node_entropy(
... [1,2,3,4,5,6,7,8,9], dx=2)
0.0
>>> global_node_entropy(
... ordinal_network([1,2,3,4,5,6,7,8,9],
... dx=2))
0.0
>>> global_node_entropy(
... np.random.uniform(size=100000), dx=3)
1.4988332319747597
>>> global_node_entropy(
... random_ordinal_network(dx=3))
1.5
\end{minted}

\section{Applications of ordinal networks with \ordpy{}}

To better illustrate the use of \ordpy{} in the context of ordinal networks, we review and replicate some literature results. Before starting, we remark that \ordpy{} does not have functions for network analysis or graph visualization. The \ordinalnetwork{} function generates output data (nodes, edges and weight lists) that can feed graph libraries such as \graphtool{}~\cite{peixoto2014graphtool}, \networkx{}~\cite{hagberg2008networkx}, and \igraph{}~\cite{csardi2006igraph}. Here, we have used \networkx{} and \igraph{}. 

\begin{figure*}[!ht]
\centering
\includegraphics[width=0.8\linewidth]{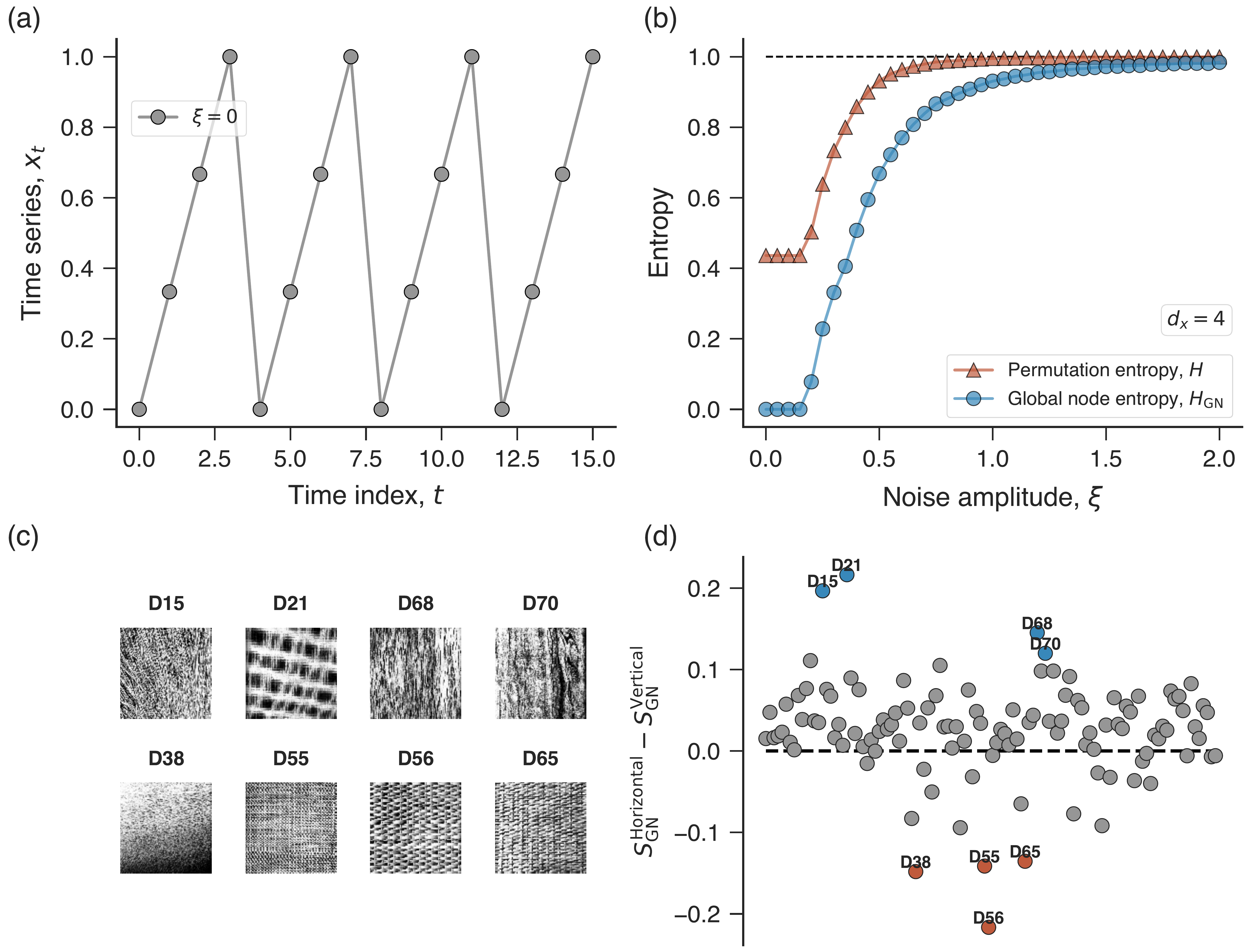}
\caption{Global node entropy of one- and two-dimensional data. (a) The initial data points of a sawtooth-like time series defined as $x_t = \{0,{1}/{3},{1}/{6},1,\dots\}$. (b) Normalized permutation entropy ($H$) and normalized global node entropy ($H_{\rm GN}$) as a function of the amplitude ($\xi$) of the uniform white noise added to the periodic sawtooth-like signals. The different curves represent average values of $H$ and $H_{\rm GN}$ over ten realizations for $\xi = \{0, 0.05,0.1,\dots,2\}$. (c) Eight examples (out of 112) of the normalized Brodatz textures. These are grayscale images (256 gray levels) with size $640\times640$~\cite{safia2020multiband}. (d) Differences between the global node entropy evaluated from the horizontal and vertical ordinal networks ($S_{\rm GN}^{\rm Horizontal} - S_{\rm GN}^{\rm Vertical}$) mapped from each Brodatz texture. We highlight eight textures (the same as shown in panel c) with the largest differences. Data and code necessary to reproduce these results are available in a Jupyter notebook at \href{http://github.com/arthurpessa/ordpy}{\ordpy{}}'s webpage.}
\label{fig:8}
\end{figure*}

We start by partially reproducing Small's~\cite{small2013complex} pioneering work in which ``ordinal partition networks'' first appeared (see Fig.~3 in that work). By following Small~\cite{small2013complex}, we numerically solve the differential equations of the R\"ossler system (with parameters $a=0.3$, $b = 2$ and $c = 4$, see Appendix~\ref{appendix:chaos} for definitions) and sample the $x$-coordinate to obtain a time series with $10^5$ observations. Figure~\ref{fig:7}{a} illustrates the periodic behavior of this time series. We then create the ordinal network from this data set with embedding parameters $d_x = 16$ and $\tau_x=1$. It is worth remembering that Small's original algorithm uses non-overlapping partitions and the edges of the resulting ordinal network are undirected and unweighted. The parameter \texttt{overlapping} in \ordinalnetwork{} should be equal to \texttt{False} to properly use Small's original algorithm. Figure~\ref{fig:7}{b} shows a visualization of this ordinal network, where the circular structure alludes to the periodicity of the original time series.

In another simple example with ordinal networks, we partially replicate Pessa and Ribeiro's~\cite{pessa2019characterizing} results on fractional Brownian motion (see Fig.~6 in their work). To do so, we generate a time series from this stochastic process with Hurst exponent $h = 0.8$ (see Appendix~\ref{appendix:stochastic} for definitions) and $2^{16}$ observations, as illustrated in Fig.~\ref{fig:7}{c}. Next, we map this time series into an ordinal network with embedding parameters $d_x = 3$ and $\tau_x = 1$ (this time using overlapping partitions as in the usual Bandt-Pompe approach). Figure~\ref{fig:7}{d} shows a visualization of the resulting ordinal network, where the persistent behavior imposed by the Hurst exponent $h = 0.8$ is captured by the quite intense autoloops associated with the ordinal patterns $(0,1,2)$ and $(2,1,0)$ (that is, the upward and downward trends of this time series). Pessa and Ribeiro~\cite{pessa2019characterizing} have also shown that local properties of ordinal networks (for instance, average weighted shortest path) are quite effective for estimating the Hurst exponent of time series, having performance superior to widely used approaches such as detrended fluctuation analysis (DFA)~\cite{peng1994mosaic}.

We also consider ordinal networks mapped from two-dimensional data. We map a periodic ornament previously explored in Ref.~\onlinecite{pessa2020mapping} (see Fig.~2 in that reference). Figure~\ref{fig:7}{e} shows the ornament of size $250 \times 250$ (see Appendix~\ref{appendix:stochastic} for more details), while Fig.~\ref{fig:7}{f} presents a visualization of the corresponding ordinal network with embedding parameters $d_x = d_y = 2$ and $\tau_x = \tau_y = 1$. 
We have made edge thickness proportional to edge weight (Eq. 22) to highlight that a few edges concentrate most of the transition probability of the network. Furthermore, we observe that this network has $12$ nodes and $72$ edges, that is, only a small fraction of all possible nodes ($24$) and edges ($416$) of a ordinal networks with $d_x = d_y = 2$ and $\tau_x = \tau_y = 1$.

In addition to the previous more qualitative examples, we have also replicated some results related to the global node entropy of ordinal networks. For time series, we follow Pessa and Ribeiro~\cite{pessa2019characterizing} (see Fig.~5 in their work) and generate a periodic sawtooth-like signal (Fig.~\ref{fig:8}{a}) with $10^5$ observations and add to it uniform white noise in the interval $[-\xi, \xi]$, where $\xi$ represents the noise amplitude. We generate these noisy sawtooth-like time series for each $\xi \in \{0,0.05,0.1,\dots,2\}$ and determine the average values of the normalized permutation entropy ($H$) and the normalized global node entropy ($H_{\rm GN}$) over ten time series replicas with $d_x = 4$ and $\tau_x = 1$. 

Figure~\ref{fig:8}{b} shows the average values of $H$ and $H_{\rm GN}$ as a function of the noise amplitude $\xi$. We note that both measures approach one with the increase of the noise amplitude. However, permutation entropy saturates for $\xi\approx1$, while global node entropy requires significantly higher values of $\xi$. This result indicates that global node entropy is more robust to noise addition and has a higher discrimination power than permutation entropy~\cite{pessa2019characterizing}.

To demonstrate the use of \globalnodeentropy{} with two-dimensional data, we calculate the global node entropy for a set of 112 8-bit images of natural textures known as the normalized Brodatz textures~\cite{safia2013new, safia2020multiband}. Figure~\ref{fig:8}{c} shows examples of these images. By following Pessa and Ribeiro~\cite{pessa2020mapping} (see Fig.~5 in their work), we calculate the global node entropy from the horizontal ($S_{\rm GN}^{\rm Horizontal}$) and vertical ($S_{\rm GN}^{\rm Vertical}$) ordinal networks mapped from the Brodatz textures with $d_x = d_y = 2$ and $\tau_x = \tau_y = 1$. 

Figure~\ref{fig:8}{d} depicts the difference between these two entropy values (that is, $S_{\rm GN}^{\rm Horizontal} - S_{\rm GN}^{\rm Vertical}$) for each Brodatz texture. We have also highlighted eight textures with extreme values for this difference. Most of these images are characterized by stripes or line segments predominantly oriented in the vertical or horizontal directions which, in turn, suggests that properties of vertical and horizontal ordinal networks can detect simple image symmetries.

\begin{figure*}[!ht]
\centering
\includegraphics[width=0.8\linewidth]{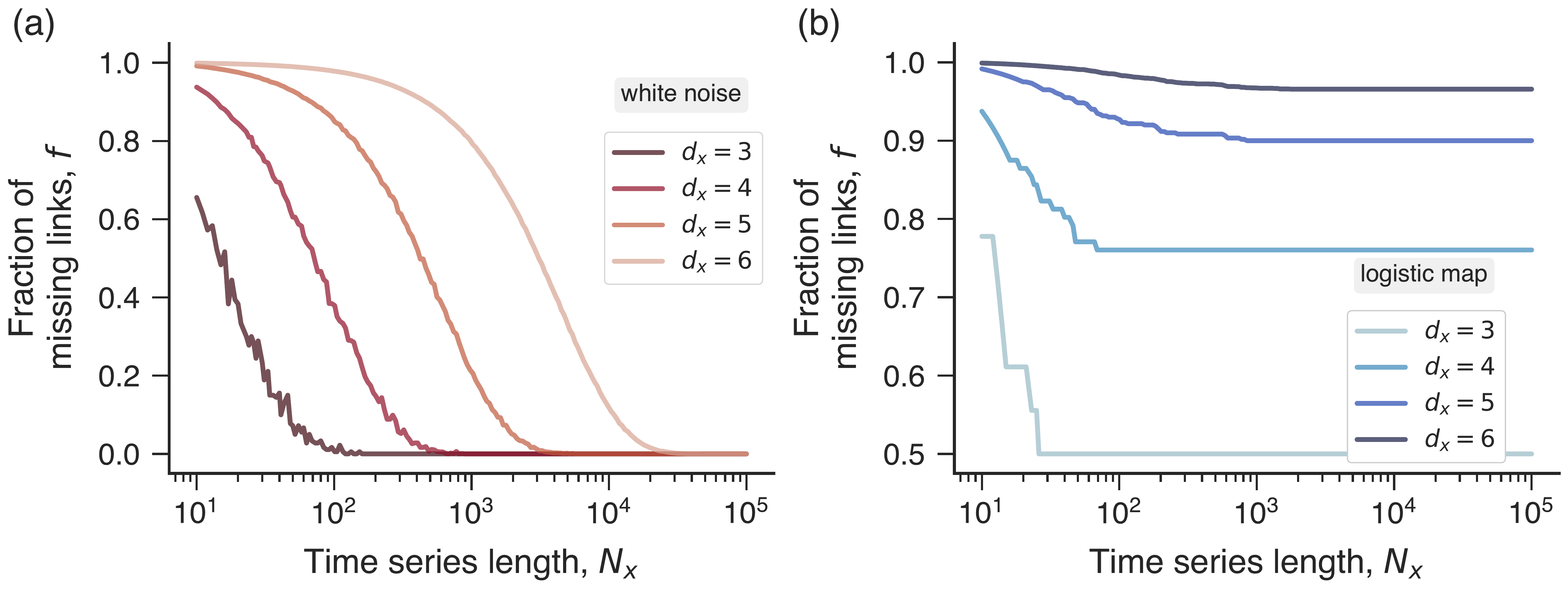}
\caption{True and false missing links in ordinal networks. (a) Dependence of the fraction of missing links ($f$) estimated from Gaussian white noise time series as a function of the time series length ($N_x$). (b) Dependence of the fraction of missing links ($f$) estimated from fully chaotic logistic time series ($r = 4$) as a function of the time series length ($N_x$). In both panels, the different curves represent average values over ten realizations (for each series length) and embedding dimension $d_x \in\{3,4,5,6\}$ with $\tau_x = 1$. We also use 194 values for $N_x$ logarithmically spaced in the interval $[10,10^5]$ in both panels. Data and code necessary to reproduce these results are available in a Jupyter notebook at \href{http://github.com/arthurpessa/ordpy}{\ordpy{}}'s webpage.}
\label{fig:9}
\end{figure*}

In a final application with ordinal networks, we explore the concept of missing links or missing transitions among ordinal patterns~\cite{pessa2019characterizing}. Similarly to the missing ordinal patterns described by Amig\'o \textit{et al.}~\cite{amigo2006order, amigo2007true}, ordinal networks can display true and false forbidden transitions among ordinal patterns. In this case, true missing links are related to the intrinsic dynamics of the process under analysis, while false missing links are associated with the finite size of empirical data sets. Because we know the exact form of random ordinal networks~\cite{pessa2019characterizing, pessa2020mapping} (here these networks represent all possible connections) we can readily find all missing links of an empirical ordinal network. In \ordpy{}, the \missinglinks{} function evaluates all missing ordinal transitions directly from a data set or the returned arrays of \ordinalnetwork{} as in: 
\begin{minted}{python}
>>> from ordpy import missing_links
>>> missing_links([4,7,9,10,6,11,3], dx=2, 
... return_fraction=False)
(array([['1|0', '1|0']], dtype='<U3'), 1)
>>> missing_links(ordinal_network(
... [4,7,9,10,6,11,3],  dx=2),
... dx=2, return_fraction=True)
(array([['1|0', '1|0']], dtype='<U3'), 0.25)
\end{minted}

To demonstrate the use of \missinglinks{} in a more engaging example, we replicate the results of Pessa and Ribeiro~\cite{pessa2019characterizing} about missing links in ordinal networks mapped from Gaussian white noise time series ( Fig.~4 in their work). We generate these time series with length $N_x$ varying logarithmically between $10$ and $10^5$, and for each one, we estimate the average fraction of missing links over ten replicas for embedding dimensions $d_x\in\{3,4,5,6\}$ and $\tau_x = 1$. Figure~\ref{fig:9}{a} shows these fractions of missing links as a function of the time series length. We observe that this quantity approaches zero as $N_x$ becomes sufficiently large. Furthermore, the smaller the embedding dimension, the faster the missing links vanish. This pattern is a fingerprint of false missing links. We have also carried out the same analysis with time series generated from logistic map iterations at fully developed chaos. Figure~\ref{fig:9}{b} shows the corresponding results. Unlike white noise, the logistic map produces ordinal networks with missing links that persist even in considerably long time series. This behavior is typical of true missing links.

\section{Conclusions}\label{sec:conclusions}

We have introduced \ordpy{} -- an open-source Python module for data analysis that implements several ordinal methods associated with the Bandt-Pompe framework. Specifically, \ordpy{} has functions implementing the following methods: permutation entropy, complexity-entropy plane, missing ordinal patterns, Tsallis and R\'enyi permutation entropies, complexity-entropy curves, ordinal networks, and missing ordinal transitions. All \ordpy{}'s functions automatically deal with one-dimensional (time series) and two-dimensional (images) data. Furthermore, most of these functions are also ready for multiscale analysis via the embedding delay parameters. Along with the description of \ordpy{} functionalities, we have also presented a literature review of several of the principal methods related to Bandt and Pompe's framework. This review further includes a reproduction of several literature results with \ordpy{}'s functions. Beyond the summarized description of \ordpy{}'s functions presented here, we notice that a complete documentation is available at \url{arthurpessa.github.io/ordpy}. All data and code used in this work are also freely available at \href{http://github.com/arthurpessa/ordpy}{\ordpy{}}'s website.

We believe \ordpy{} will help to popularize ordinal methods even further, particularly in research fields with more limited tradition in scientific computing. In addition to a myriad of possible empirical applications, we also believe \ordpy{} can further promote the development of new methods related to Bandt and Pompe's framework. We remark that some techniques available in \ordpy{} have received little attention or have not even been formally proposed. These possible developments already implemented in \ordpy{} include the use of complexity-entropy curves for two-dimensional data, multiscale complexity-entropy curves, ordinal networks with different embedding delays (particularly for two-dimensional data), analysis of missing patterns in two-dimensional data, and missing ordinal transitions. We also plan to implement more techniques based on the Bandt and Pompe's framework and include them in future versions of \ordpy{}.

Finally, we hope our module helps making research methods more accessible and reproducible~\cite{baker2016interactive, fanelli2018opinion} as well as other open-source software efforts such as the \tisean{}~\cite{hegger1999tisean} (nonlinear time series analysis), \pyunicorn{}~\cite{donges2015unified} (time series networks and recurrence analysis), and \powerlaw{}~\cite{alstott2014powerlaw} (analysis of heavy-tailed distributions) packages.

\begin{acknowledgments}
This research was supported by Coordena\c{c}\~ao de Aperfeicoamento de Pessoal de N\'ivel Superior (CAPES) and Conselho Nacional de Desenvolvimento Cient\'ifico e Tecnol\'ogico (CNPq -- Grants 407690/2018-2 and 303121/2018-1).
\end{acknowledgments}

\section*{Data Availability}
All data and code necessary to reproduce the results and figures of this work are available at \url{http://github.com/arthurpessa/ordpy}, Ref.~\onlinecite{pessa2021ordpygithub}.

\appendix
\section{Selection of embedding parameters}\label{appendix:parameter}

As we have commented in the main text, the embedding parameters ($d_x$ and $\tau_x$) are important for several applications related to the Bandt-Pompe framework, and wrong choices can lead to misleading conclusions. At the same time, there is no unique fail-safe procedure for selecting optimal values for these parameters, and this choice often depends on the time-series nature and the research question under analysis. In the context of permutation entropy, Myers \textit{et al.}~\cite{myers2020automatic} suggest three main strategies: \textit{i)} follow experts' suggestion; \textit{ii)} trial and error; and \textit{iii)} the use of nonlinear time series methods related to phase space reconstruction.

The first strategy consists of following good practices previously established in the literature, and good starting points are review articles on permutation entropy and related methods such as Refs.~\onlinecite{zanin2012permutation, riedl2013practical, amigo2015ordinal, keller2017permutation}. The work of Riedl \textit{et al.} is particularly interesting for this strategy as the authors compile different choices of embedding parameters according to characteristics of time series and research field. Among other propositions, these authors suggest using $\tau_x = 1$ and the largest embedding dimension yielding a proper evaluation of the ordinal distribution when dealing with data more easily described by discrete models~\cite{riedl2013practical}.

The second strategy refers to the computational origin of the Bandt-Pompe framework, and much in line with statistical learning methods~\cite{james2014introduction, gueron2017handson}, optimal parameter selection is often achieved by experimentation (trial and error), heuristics, and validation using null models. We believe this strategy is fundamental in applications involving classification and regression tasks, where the optimal embedding parameters can be found by optimizing loss functions in cross-validation and train/test split strategies~\cite{cuesta2018patterns}. For instance, Kulp~\textit{et al.}~\cite{kulp2016using-ordinal} have suggested using ensemble of random series with the same length of the series under analysis and selecting the maximum embedding dimension for which the number of missing patterns is zero.

The third strategy for selecting the embedding parameters refers to using methods derived or related to nonlinear time series analysis~\cite{kantz2004nonlinear, bradley2015nonlinear, small2005applied}. Common techniques such as looking for the fist zero of the autocorrelation function or the first minimum of the mutual information might be especially interesting when choosing $\tau_x$~\cite{kantz2004nonlinear, small2005applied}. It is worth remembering that the concept of embedding parameters in the Bandt-Pompe approach is intimately related to the idea of embedding and phase-space reconstruction in the context of dynamical systems~\cite{packard1980geometry, kennel1992determining, cao1997practical, fraser1986independent}. Indeed, investigations based on ordinal methods in the context of chaotic dynamics are an instrumental part for the development of the Bandt-Pompe framework~\cite{bandt2002permutation, rosso2007distinguishing, small2013complex}. In this context, a simple and interesting conceptualization on how the embedding dimension relates to the underlying phase space is presented by Groth~\cite{groth2005visualization}.

\begin{figure*}[!ht]
\centering
\includegraphics[width=.8\linewidth]{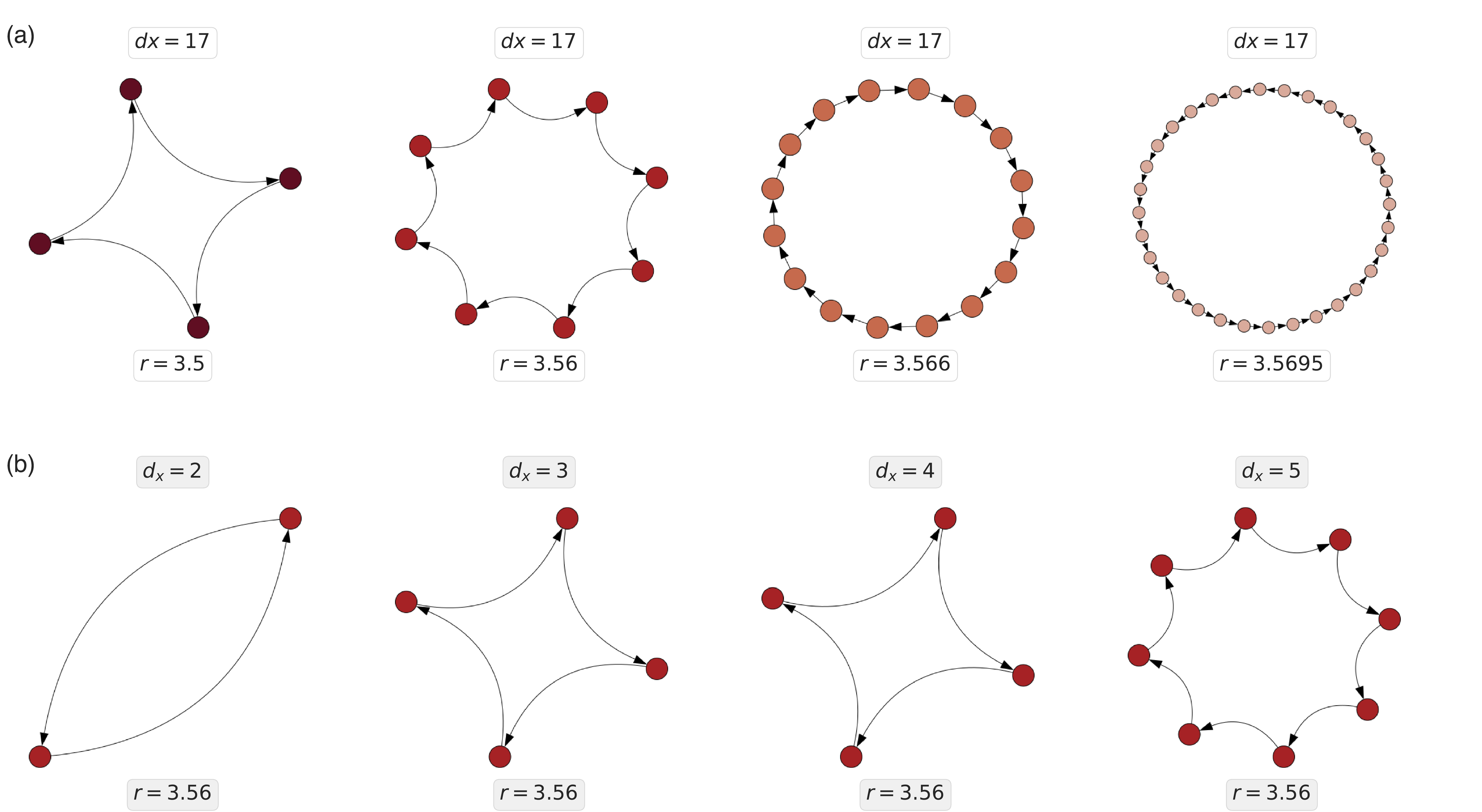}
\caption{
Ordinal networks mapped from periodic logistic series. (a) Ordinal networks mapped from periodic logistic series $(N_x = 100)$ with parameters $r = 3.5$ (period 4), $r = 3.56$ (period 8), $r = 3.566$ (period 16), and $r = 3.5695$ (period 32). All four networks are obtained with embedding parameters $d_x = 17$ and $\tau_x = 1$. (b) Ordinal networks mapped from a logistic series of period 8 ($r = 3.56$) with embedding dimensions $d_x \in \{2, 3, 4, 5\}$ and $\tau_x = 1$. We notice that $d_x$ must be larger than $4$ so that the number of nodes is equal to the time series period. We further remark that this network structure is not modified when considering values $d_x$ larger than $4$ (as shown by the second network of panel a). Data and code necessary to reproduce these results are available in a Jupyter notebook at \href{http://github.com/arthurpessa/ordpy}{\ordpy{}}'s webpage.
}
\label{fig:10}
\end{figure*}

In addition to the previous three main strategies, there are also attempts devoted to developing automatic procedures for selecting embedding parameters in the context of permutation entropy.~\cite{myers2020automatic, wang2019optimized, riedl2013practical}. Among these, we highlight the interesting comparison between expert recommendations and automatic approaches presented by Myers \textit{et al.}~\cite{myers2020automatic}. 

Considering the more recent developments related to mapping time series into ordinal networks, several works on this topic have devoted efforts to the optimal selection of embedding parameters~\cite{small2013complex, mccullough2015time, mccullough2017regenerating, sakellariou2019markov}. In this context, topological properties or network metrics have become major criteria for properly selecting embedding parameters capable of capturing dynamical features of time series. To illustrate this approach, we have partially reproduced the results of Sakellariou \textit{et al.}~\cite{sakellariou2019markov} (see Fig.~8 in their work) about ordinal networks mapped from periodic logistic series with period $2^k$ for $k\in(2,3,4,5)$. Figure~\ref{fig:10}{a} shows a representation of these networks mapped with $d_x = 17$ and $\tau_x=1$. We observe that this value of $d_x$ is large enough to map the periodic behavior of these time series into a regular ring-like network structure with the number of nodes precisely equal to the time series period. As discussed by Sakellariou \textit{et al.}~\cite{sakellariou2019markov}, the embedding dimension needs to be larger than $2^{k-1}$ so that the network topology explicitly represents the period $2^k$ of the time series. Figure~\ref{fig:10}{b} illustrates what happens with ordinal networks mapped from a period 8 time series ($k=3$) for different values of $d_x$, confirming the network topology only explicitly accounts for period 8 behavior for $d_x>4$. Similar problems involving non-optimal choices for $d_x$ emerge when using ordinal networks to estimate dynamical quantities such as the topological entropy~\cite{sakellariou2021estimating}.

\section{Definitions of dynamical systems}\label{appendix:chaos}
In this appendix, we present a brief definition of the dynamical systems used in this manuscript.

\begin{enumerate}[leftmargin=0cm,itemindent=.5cm,labelwidth=\itemindent,labelsep=0cm,align=left]
\item The logistic map is defined by the following difference equation~\cite{may1976chaos}:
\begin{equation}\label{eq:logistic_map}
    x_{t+1} = r x_t (1 - x_t)\,,
\end{equation}
where $r$ is a parameter. We have used $r = 4$ in most applications of this manuscript unless specified otherwise.

\item The transient logistic map is defined as~\cite{trulla1996recurrence, cao2004detecting}
\begin{equation}
    x_{t+1} = r(t) x_t (1 - x_t)\,,
\end{equation}
where the parameter $r(t)$ changes at each iteration. The results of Figs.~\ref{fig:3}{e} and \ref{fig:3}{f} were obtained with $r(t=0)=3.5$ and by incrementing this parameter in steps of size $10^{-4}$ up to $r(t)=4$.

\item The skew tent map is defined as~\cite{sakai1980autocorrelations}
\begin{equation}
\begin{cases}
   x/\omega \hspace{2.05cm} & \text{for } x \in [0,\omega] \\
   (1-x)/(1-\omega) & \text{for } x \in [\omega, 1]
\end{cases}\,,
\end{equation}
where $\omega$ is a parameter. In the complexity-entropy plane application shown in Fig.~\ref{fig:4}{a}, we have used $\omega=0.1847$.

\item The H\'enon map is defined as~\cite{schuster2006deterministic}
\begin{equation}
\begin{cases}
    x_{t+1} = 1 - a x_t^2 + y_t & \\
    y_{t+1} = b x_t &
\end{cases}\,,
\end{equation}
where $a$ and $|b| < 1$ are parameters. This map can be thought of as a two-dimensional extension of the logistic map~\cite{schuster2006deterministic}. We have used $a=1.4$ and $b=0.3$ for the results related to the complexity-entropy plane.

\item The Schuster map is defined as~\cite{schuster2006deterministic}
\begin{equation}
    x_{t+1} = (x_t + x_t^z)\text{ mod 1}\,,
\end{equation}
where $z$ is a parameter, and the modulo operation returns the fractional (decimal) part of a number. We have used $z \in \{2,2.5,3\}$, as shown in Fig.~\ref{fig:4}{a}.

\item The R\"ossler system is a continuous time dynamical system defined as~\cite{rossler1976equation, strogatz2014nonlinear}
\begin{equation}\label{eq:rossler}
\begin{split}
    \frac{dx}{dt} &= - y - z \\
    \frac{dy}{dt} &= x + ay \\
    \frac{dz}{dt} &= b + z(x - c)
\end{split}\,,
\end{equation}
where $a$, $b$, and $c$ are parameters. We have numerically solved this differential equation system using the \scipy{} Python module~\cite{scipy} with parameters $a=0.3$, $b = 2$ and $c = 4$.

\end{enumerate}

\section{Definitions of stochastic processes}\label{appendix:stochastic}
In this appendix, we briefly describe the stochastic processes used in the manuscript.

\begin{enumerate}[leftmargin=0cm,itemindent=.5cm,labelwidth=\itemindent,labelsep=0cm,align=left]
\item An Ising surface~\cite{brito2007dynamics, brito2010two} is a square lattice in which the height at each lattice site represents the accumulated sum of spin variables of particles in a Monte Carlo simulation~\cite{landau2015guide}. If we assume $\sigma_i \in \{-1,1\}$ represents the spin variable at site $i$, we can write the Hamiltonian of this system as
\begin{equation}
    \mathcal{H} = -\sum_{\langle i,j \rangle}\sigma_i \sigma_j\,,
\end{equation}
where the summation is over all pairs of first neighbors in a square lattice. The height $S_i$ at site $i$ of the corresponding Ising surface is then defined as
\begin{equation}
    S_i = \sum_{t} \sigma_i(t)\,,
\end{equation}
where $\sigma_i(t)$ is the spin value in step $t$ of the Monte Carlo simulation. For each surface, we define the reduced temperature $T_r$ as the ratio between the temperature $T$ and the critical temperature $T_c$ of the Ising system $(T_c = 2/\ln{(1+\sqrt{2})})$. Finally, we have used periodic boundary conditions in our numerical experiments.

\item A fractional Brownian motion is a continuous, self-similar, and non-stationary stochastic process introduced by Mandelbrot and Van Ness~\cite{mandelbrot1968fractional}. The Hurst exponent $h \in (0,1)$ controls the roughness observed in samples of this process, such that the smaller the values of $h$, the rougher the time series. The case $h = 1/2$ corresponds to ordinary Brownian motion (integrated Gaussian white noise). To generate samples (time series) of this stochastic process, we have used the Hosking method~\cite{hosking1984modeling}. 

\item A fractional Gaussian noise is a stationary stochastic
process that represents the increments of fractional Brownian motion. For this Gaussian process, the Hurst parameter $h \in (0,1)$ controls the range of auto-correlation of the time series. For $h > 1/2$, the process presents long-range persistent memory. For $h < 1/2$, samples present anti-persistent behavior. We have Gaussian white noise if $h =1/2$. To generate samples of a fractional Gaussian noise, we have also used the Hosking method~\cite{hosking1984modeling, diecker2004simulation}. More detailed information about simulations of fractional Gaussian noise and fractional Brownian motion can be found in Ref.~\onlinecite{diecker2004simulation}. The C source code used in this work is publicly available in Ref.~\onlinecite{diecker2020hosking}.

\item A $1/f$ noise or a Flicker noise~\cite{voss1979flicker, kasdin1995discrete} is a class of stochastic processes presenting a power-law power spectral density~\cite{voss1979flicker, timmer1995generating, kasdin1995discrete}, that is, $\mathcal{P} \sim 1/f^{-k}$. The case $k = 0$ corresponds to white noise, while $k = 2$ corresponds to brown noise (random walk or integrated white noise). We have generated Gaussian distributed $1/f^{-k}$ noise for $k \in \{0, 0.25, 0.50, \dots , 3.0\}$ with the algorithm proposed by Timmer and K\"onig~\cite{timmer1995generating} as implemented in Ref.~\cite{patzelt2020colored}.

\item The periodic ornament used in this work can be generated by first defining two square arrays
\begin{equation}
    X_{i, j} = \frac{2\pi(j - 1)}{(n-1)} ~\text{ and }~ Y_{i, j} = \frac{2\pi(i - 1)}{(n-1)}\,,
\end{equation}
and next by calculating
\begin{equation}
    Z_{i, j} = \sin\left(\frac{\omega}{2\pi} X_{i,j}\cos{\theta}  - \frac{\omega}{2\pi} Y_{i,j}\sin{\theta}\right)\,,
\end{equation}
where $i, j = 1,\dots,n$, with $n$ being the ornament size, $\theta$ defining the stripes angle, and $\omega > 0$ the stripe frequency. The ornament shown in Fig.~\ref{fig:7}{e} is obtained by setting $n = 250$, $\omega = 9$ and $\theta = 135$ degrees. Previous works~\cite{zunino2016discriminating, brazhe2018shearlet, pessa2020mapping} have also considered shuffled versions of this periodic ornament, where a parameter controls the fraction of elements $Z_{i,j}$ that are randomly shuffled. A function implementing this geometric ornament is available in \ordpy{}'s notebook.
\end{enumerate}

\bibliography{ordpy}

\end{document}